\documentclass[letterpaper]{article} 
\usepackage{aaai2026}  
\usepackage{times}  
\usepackage{helvet}  
\usepackage{courier}  
\usepackage[hyphens]{url}  
\usepackage{graphicx} 
\usepackage{natbib}  
\usepackage{caption} 
\usepackage[ruled,vlined]{algorithm2e}
\usepackage{multirow}
\usepackage{amsmath}
\usepackage[htt]{hyphenat}
\usepackage{ragged2e}
\usepackage{lineno}

\usepackage{tcolorbox}
\tcbuselibrary{listings}
\usepackage{pbox}

\newtcblisting{promptbox}[1]{%
  title=#1,
  listing only,
  colback=white!95!gray,
  colframe=gray,
  width=\columnwidth,
  arc=0mm,
  boxrule=0.5mm,
  colbacktitle=white!95!gray,
  coltitle=black,
  fonttitle=\bfseries,
  left=6pt,
  right=6pt,
  top=6pt,
  bottom=6pt,
  listing options={basicstyle=\small\ttfamily, breaklines=true}
}

\tcbset{
  prompt/.style={
    colframe=gray!60,
    boxrule=0.5pt
    }
}

\usepackage{newfloat}
\usepackage{listings}
\DeclareCaptionStyle{ruled}{labelfont=normalfont,labelsep=colon,strut=off} 
\title{Simulating Online Social Media Conversations on Controversial Topics Using AI Agents Calibrated on Real-World Data}

\author{
    Elisa Composta\textsuperscript{\rm 1},
    Nicolò Fontana\textsuperscript{\rm 1}\equalcontrib,
    Francesco Corso\textsuperscript{\rm 1, \rm2}\equalcontrib,
    Francesco Pierri\textsuperscript{\rm 1}
}
\affiliations{
    \textsuperscript{\rm 1}DEIB, Politecnico di Milano, Via Golgi 34, Milano, Italy

    \textsuperscript{\rm 2}CENTAI, Corso Inghilterra 4, Torino, Italy \\
     elisa.composta@mail.polimi.it, nicolo.fontana@polimi.it, francesco.corso@polimi.it, francesco.pierri@polimi.it\\
    \noindent\textcolor{red}{\textbf{Please check the published version of the paper at AAAI ICWSM 2026}}
}

\usepackage{enumitem}
\usepackage{xcolor}
\newcommand{\answerYes}[1]{\textcolor{blue}{#1}} 
 
\newcommand{\answerNA}[1]{\textcolor{gray}{#1}}

\newcommand{\bea}{\begin{eqnarray}} 
\newcommand{\eea}{\end{eqnarray}}

\newcommand{\arti}[1]{$\texttt{artifish/llama3.2-uncensored-3b}$}
\newcommand{\artiname}[1]{$\texttt{Llama3.2-3B}$}
\newcommand{\llname}[1]{$\texttt{Llama2-70B}$} 

\begin{document}

\maketitle

\begin{abstract}
Online social networks offer a valuable lens to analyze both individual and collective phenomena.
Researchers often use simulators to explore controlled scenarios, and the integration of Large Language Models (LLMs) makes these simulations more realistic by enabling agents to understand and generate natural language content. In this work, we offer an exploratory investigation of the behavior of LLM-based agents in a simulated microblogging social network.
We initialize agents with realistic profiles and opinions calibrated on real-world online conversations from the 2022 Italian political election and extend an existing simulator by introducing mechanisms for opinion modeling.
We examine how LLM agents simulate online conversations, interact with others, and evolve their opinions under different scenarios.
Our results show that LLM agents generate coherent content, form connections, and build a realistic social network structure.
However, their generated content displays less heterogeneity in tone and toxicity compared to real data.
We also find that LLM-based opinion dynamics evolve over time in ways similar to traditional mathematical models.
Varying parameter configurations produces no significant changes, indicating that simulations require more careful cognitive modeling at initialization to replicate human behavior more faithfully.
Overall, we demonstrate the potential of LLMs for simulating user behavior in social environments, while also identifying key challenges in capturing heterogeneity and complex dynamics.

\end{abstract}

\section{Introduction} \label{sec:introduction}

Online social networks have evolved beyond mere communication platforms, becoming digital arenas where users express emotions, form opinions, and shape behaviors~\cite{bakshy2015}. 
These environments provide unique opportunities to study complex collective phenomena such as polarization, content diffusion, and large-scale social dynamics~\cite{vosoughi2018spread}. 
The widespread availability of digital traces, enabled by the pervasive use of online technologies, has fueled the rise of computational social science~\cite{lazer2009computational}, which seeks to explain human behavior and social processes through computational methods.

A prominent approach in this field involves the use of simulation tools~\cite{squazzoni2014socialsimulation}. 
Simulations make it possible to create controlled virtual environments where researchers can test hypotheses, compare strategies, and observe the evolution of user behavior under conditions that would be difficult—or ethically problematic—to reproduce in the real world~\cite{rossetti2024ysocialllmpoweredsocial}. 
For example, one can investigate the spread of harmful content and rumors~\cite{hu2025simulatingrumorspreadingsocial} or assess how different recommendation algorithms shape user activity~\cite{törnberg2023evaluate}, all without intervening directly on live platforms.
Despite their promise, building realistic simulations of online social networks remains challenging. 
Emergent behaviors in these systems are driven by numerous individual-level factors that are difficult to predict or formalize. 
Human interactions involve ambiguity, context-dependence, and variability, which complicate the design of models capable of capturing social complexity~\cite{gao2023s3socialnetworksimulationlarge}.
Agent-Based Modeling (ABM) has long been employed to address this challenge. ABMs represent systems as collections of autonomous agents, each following a set of predefined and simplified behavioral rules~\cite{macy2002abm, conte2014agent}. 
While such models have yielded important insights, they struggle to capture the richness of human behavior, which is mediated by language, emotions, and social context~\cite{törnberg2023evaluate}.

In this context, Large Language Models (LLMs) represent a promising extension of ABM.
Unlike traditional agents, which follow fixed and narrow rules, LLM-based agents can generate nuanced, coherent, and context-aware behaviors~\cite{park2024generative,fontana2025nicer}. 
Their capacity to simulate conversations, express emotions, adopt perspectives, and deploy diverse interaction strategies enables them to approximate human-like behavior with unprecedented fidelity~\cite{park2023genagents,corso2025androids}. 
Moreover, LLM agents can be enriched with persistent traits—such as personality, ideology, or memory of past interactions—that allow them to act consistently across time~\cite{rossetti2024ysocialllmpoweredsocial}. 
This makes them particularly powerful for reproducing both individual-level realism and emergent collective phenomena~\cite{park2023genagents}.
Early studies have demonstrated the potential of LLM-driven simulations, suggesting that these models offer a promising avenue for advancing the study of online behavior and warrant systematic exploration~\cite{gao2023s3socialnetworksimulationlarge, törnberg2023evaluate, rossetti2024ysocialllmpoweredsocial}.

This work builds on \textit{Y Social}~\cite{rossetti2024ysocialllmpoweredsocial}, a social media simulator that reproduces online platforms within a controlled environment. 
In this simulator, LLM-powered agents behave like users: they consume content, post, and interact with one another.
Using a data-driven approach that leverages Y Social framework with real-world online conversations from the 2022 Italian election, we aim to address the following research questions:
\begin{itemize}
    \item \textbf{RQ1:} How realistically do LLM-based agents reproduce in-/out-group dynamics among supporters of different political parties?
    \item \textbf{RQ2:} How do opinion dynamics generated by LLM agents differ from those predicted by traditional mathematical models?
\end{itemize}
To this end, our contribution extends Y in two key directions.
First, we seed agents with activity patterns and political leaning drawn from a dataset of real-world Twitter conversations collected around the 2022 Italian elections~\cite{pierri2023ita}, enhancing the realism of their behavior and enabling systematic comparison with an empirically grounded scenario. Second, we equip agents with opinions on specific political topics and observe the dynamics of opinion change over time through interactions.
We conduct a series of experiments varying key parameters such as the LLM powering agents, the network structure, and the recommendation system. 
We then compare simulated opinion dynamics outcomes against real-world data to assess both the validity and the limitations of LLM-based social simulations.
By doing so, this study both explores the potential of LLMs as social agents and identifies the challenges that must be addressed to use them as reliable tools for modeling social behavior.

\section{Related Work}\label{sec:related_work}

\subsection{Simulating social networks}

Social simulations have been widely used to study group behavior and the nonlinear effects of individual interactions~\cite{squazzoni2014socialsimulation}. Agent-Based Modeling (ABM) focuses on local agent dynamics and demonstrates how simple interactions can reproduce complex social phenomena~\cite{macy2002abm}. ABMs also allow the linking of micro- and macro-level phenomena, highlighting causal relationships between individual behavior and network structure~\cite{squazzoni2014socialsimulation}.
A key limitation of traditional ABMs is the simplicity of agent behavior rules~\cite{conte2014agent} and their limited capacity for realistic social interaction~\cite{törnberg2023evaluate}. Recent advances in AI and Large Language Models (LLMs) offer a way to overcome these constraints, enabling agents to engage in realistic conversations and exhibit human-like behaviors~\cite{park2023genagents}.
Several recent studies have explored the use of LLM agents in simulated social network environments. Below, we discuss three representative simulators.

\citet{törnberg2023evaluate} simulated three social media platforms, each with a distinct content recommendation algorithm, to assess how news feed personalization affects conversation quality and cross-party interactions. 
Agents, powered by LLMs, were initialized with demographic characteristics, political leanings, interests, and attitudes from the 2020 American National Election Study (ANES).
The first platform promoted popular posts from followed users, while the second suggested globally popular posts; both reduced cross-party interactions and increased toxicity.
In contrast, a third ``bridging'' algorithm recommended posts popular among users with opposing views, resulting in more constructive, less toxic, and more inter-partisan interactions.
This work underscores the significant influence of recommender systems on online discourse quality.


\citet{gao2023s3socialnetworksimulationlarge} proposed the $S^3$ system, in which LLM agents maintain a memory pool of their most relevant posts.
This allows agents to preserve cognitive coherence and realism over time, with decisions influenced by past actions rather than treated independently, mimicking real-world user behavior.
$S^3$ was evaluated on real-world social network data at both individual and population levels. 
At the individual level, the study examined emotions, attitudes, and content generation; at the population level, it assessed information propagation and the spread of emotions and attitudes.
Results indicate that the system accurately replicates complex dynamics observed in real networks, demonstrating that memory-equipped LLM agents can provide realistic insights at both micro- and macro-levels.

\citet{rossetti2024ysocialllmpoweredsocial} introduced Y, a social media digital twin, a system designed to digitally replicate a real-world system to allow analysis, simulation, and experimentation in a controlled environment.
The users of these simulations are LLM agents, and they can perform all the common actions available on the most popular social media, including posting, commenting, replying, reacting, and following other users. Other modules also allow the integration of images.
User profiles are enriched with attributes including their interests, political leaning, demographic data, and personality, which is defined according to the Big Five model~\cite{barrick1991bigfive, McCrae1992}.
To make the simulations even more realistic, Y also includes the possibility of adding external input to the simulation. Specifically, users can share news gathered from selected websites, provided through RSS (Really Simple Syndication) feeds.
Moreover, Y includes the implementation of various recommender and ranking algorithms to promote specific content or users. This enables further study of the impact the algorithmic curation has on online conversations and users' behavior. This expands the approach of~\citet{törnberg2023evaluate}by offering a more flexible and realistic simulation framework.

\subsection{Opinion Dynamics}
Modeling opinion dynamics (i.e., how individuals update their views through interaction) has traditionally relied on mathematical abstractions. 
A classic example is the DeGroot model~\cite{Degroot1974}, where each opinion is updated as a weighted average of neighbors’ opinions.
While capturing social influence, this model assumes full susceptibility and ignores resistance to change. 

The Friedkin–Johnsen model~\cite{friedkin_1990} addresses this by introducing a susceptibility parameter, allowing agents to retain part of their initial opinion. 
More recent extensions incorporate state-dependent updating, where adjustments depend on current beliefs rather than the initial stance~\cite{Ye2018Opinion, Liu_2018}.

While mathematical models provide valuable abstractions, they reduce opinions to numerical values and overlook elements such as language, tone, and personality. 
Recent studies address this by employing LLMs as agents, which can impersonate profiles, engage in realistic interactions, and express beliefs in natural language.
For example,~\citet{cau2025languagedrivenopiniondynamicsagentbased} simulated paired discussions on the Ship of Theseus paradox, a topic without factual resolution. 
Agents held discrete opinions (0–6) and updated them stepwise if persuaded. 
Results showed that LLMs often aligned with partners’ views, but the setup underutilized LLM capabilities, as agents lacked richer personalization, such as demographics or personality traits.

\citet{gao2023s3socialnetworksimulationlarge} modeled attitude evolution as a Markov process on a binary spectrum, where LLM agents initialized with predefined profiles update their belief states by evaluating incoming messages.
Similarly,~\citet{chuang2024simulatingopiniondynamicsnetworks} studied dyadic interactions, mapping textual replies to numerical scores through a classifier.
Their findings show that while LLMs tend to converge toward accurate information, replicating human behavior requires introducing confirmation bias, as real users often reinforce prior beliefs.

\citet{Liu_2024} simulated social media with LLM agents modeled as detailed personas with memory modules, expressing opinions in tweets and updating them after random exposures.
While capturing dynamic content, the setup lacked a realistic social graph, as propagation ignored network structure and recommendation effects.
Building on this,~\citet{piao2025emergencehumanlikepolarizationlarge} showed that LLMs can mirror human patterns: converging on fact-based topics (e.g., flat Earth) while polarizing on political issues. 
Their framework mapped agents’ self-rated political leanings to numerical scores, aligning language with opinion measures.

\section{Experimental Design} \label{sec:setup} 


To conduct our simulations, we employed Y~\cite{rossetti2024ysocialllmpoweredsocial}, a social media twin framework designed to replicate user interactions and dynamics on platforms structured similarly to X (formerly Twitter).
We selected this framework due to its modular architecture and its realistic temporal activity model, which has been fitted on Bluesky Social data, as well as the native support for different configurable recommender algorithms, new user generation, and LLM-powered agents~\cite{rossetti2024ysocialllmpoweredsocial, failla2024}.
In our simulations, each agent was simulated by an LLM.
Specifically, we employed Ollama to run uncensored versions of 
\texttt{Llama2} (70B parameters\footnote{https://ollama.com/library/llama2-uncensored:70b}) and \texttt{Llama3.2} (3B parameters\footnote{https://ollama.com/artifish/llama3.2-uncensored:latest}), which enabled discussions of controversial topics without triggering safety-filter refusals.
All experiments were conducted with a temperature setting of 0.9, chosen to encourage response diversity while minimizing hallucinations and nonsensical outputs.
The prompt used contained the most relevant information needed by the agents to contextually make reasonable choices:
\begin{itemize}
    \item \textbf{Demographics.}  
    Agents are assigned the following attributes: \texttt{age} (integer, sampled from weighted 2024 X statistics, range 18--60), \texttt{gender} (categorical; binary variable, sampled with probabilities proportional to the gender distribution observed on the real platform based on Statista~\cite{statista2024twitter}), and \texttt{nationality} (fixed to Italian to match the case study).  
    These attributes are used as input features to the prompts and affect probabilities for actions such as posting, commenting, or following, but also unfollowing and, more in general, all the possible interactions that the agents can perform, together with the internal opinion-updating mechanism.
    We leave more details in Appendix \ref{app:prompt_actions}.
    \item \textbf{Political leaning \& coalition principles.}  
    Each agent receives a \texttt{coalition} label (i.e., Right, Centre-Left, Third Pole, M5S) sampled from the source dataset and inferred based on the retweeting behaviour with respect to representatives of different parties~\cite{pierri2023ita}. For each coalition, we store a short, standardized \texttt{principles} string summarizing its typical policy priorities and rhetorical framing (used as contextual prompt material for the LLM). Political leaning is represented categorically (\texttt{coalition}), whereas topic opinions are represented on a continuous range to seed initial opinion values. Additional information on the coalitions can be found in Appendix \ref{app:coalition_opinions}.
    
    \item \textbf{Current opinions.}  
    For every topic, agents maintain a \texttt{stance\_score} (numeric) and a \texttt{justification} (short text). The numeric score is mapped to the interval \([-1,+1]\) (for example: strongly oppose \(\sim -1\), neutral \(=0\), strongly support \(\sim +1\)). The \texttt{justification} is a brief textual rationale, describing why the agent holds that position; it is both an output of the opinion-update process and a contextual input to subsequent LLM-generated utterances, ensuring consistency between numeric and linguistic representations.
    
    \item \textbf{Topic descriptions.}  
    Each topic has a canonical \texttt{description} and explicit definitions of what constitutes \texttt{supportive} vs \texttt{opposed} stances. These are used to (i) disambiguate labels for the LLM prompts and (ii) map textual judgments to numeric scores.
\end{itemize}





To make our agents more realistic, some attributes have been initialized based on the ITA-ELECTION-22 dataset~\cite{pierri2023ita}, a collection of Twitter posts in the Italian language around the Italian political election in 2022.
Specifically, the attributes initialized from the dataset in this work are: the political leaning, the average toxicity of posts and comments posted on the platform (estimated using Detoxify), and the activity level, for each user.
The activity is computed by converting the number of tweets posted by each user into a continuous value in the range $[0,1]$, with a logarithmic normalization to reduce the impact of outliers.
The formula used is the following:

\[
activity_x = \min\left( \frac{\log(1 + n\_posts_x)}{\log(1 + N_{99.5})},\ 1.0 \right)
\]

where $n\_posts_x$ is the number of posts written by user $x$, and $N_{99.5}$ is the 99.5th percentile.

\subsubsection{Network of interactions and simulation steps}
We adopt two network initialization strategies.
\begin{itemize}
    \item \textbf{Empty Network}: the social network is empty: there are no connections between the nodes.
    \item \textbf{Fully connected Network}: every node (user) in the social network is connected to every other node.
\end{itemize}
These two initialization strategies represent the two extreme regimes of social connectivity: the Empty Network captures a scenario with no social influence, where users’ behavior is driven solely by individual cognition and content exposure, while the Fully Connected Network represents the maximal influence regime, where every user can directly observe and react to every other user. 
Together, these extremes represent the upper and lower bound the spectrum of possible interaction structures, allowing us to isolate the effect of social connectivity itself on emergent dynamics.
The agents create the network connections over time via the \texttt{follow} interaction, allowing the network structure to emerge dynamically as the simulation progresses.
In the extended framework, the opinion update is directly performed by LLMs.
The evolving network structure was then used to compute the mathematical equivalent of agents’ stances through the Friedkin–Johnsen (FJ) model, which we executed in parallel to our simulation, in order to compare the results of these two approaches. 
We chose to rely on the FJ model as our baseline since it is widely used in Agent-based modeling literature~\cite{sun2023opinion,disaro2024balancing} and, even in its original version, is easily comparable to our LLM-based setup.

At the beginning of the simulation, a population of agents is initially generated.
At this stage, the agents may already be connected with each other, depending on the network initialization strategy.
However, throughout the simulation, agents have the possibility to create new links or remove the existing ones, evaluating the interactions they had with other users.
In this way, the network structure dynamically evolves over time according to the agents' behavior and interactions.
In Procedure 1, we then describe each step of a day in the simulation.

Each simulated day is composed of a set of rounds, corresponding to virtual hours.
In each round, a number of active agents is sampled, according to the hourly activity configured.
Agents can then perform an action: publish content, react, follow or unfollow other users, and eventually reply to previously received mentions.
The specific behavior of each agent depends on its profile, its personality and the content it's interacting with. 

At the end of the day, active agents are asked to update their opinion on the topics they discussed.
This phase is critical to study the opinion dynamics: it makes it possible to observe how social interactions and the received content impact the evolution of individual views.
We normalized all numeric attributes (facilitating their use in probabilistic decision functions), and we stored textual fields (principles, justifications, topic descriptions)  as compact prompt templates to ensure consistent LLM conditioning.
The behavioral \texttt{Decision(profile, content)} function combines numeric traits, coalition priors, and content-topic alignment to produce an action probability distribution.

\subsubsection{Topics of discussion}
The discussions between agents focused on four major topics of debate that characterized the Italian 2022 election, selected for their political salience and the presence of a range of stances for each of these issues.
These topics were selected because they represent politically relevant issues in the Italian context of 2022, on which the main coalitions held different positions.
This allows the simulation to generate meaningful political discussions and potential conflicts among agents.
Furthermore, since these issues are characterized by many different stances, the simulations do not necessarily lead to consensus~\cite{cau2025languagedrivenopiniondynamicsagentbased}.
\begin{itemize}
    \item \textbf{Civil rights:} covering gender equality, LGBTQIA+ rights, and family structure.  
    \item \textbf{Immigration:} centered on border control, bilateral agreements, and the management of irregular migration.  
    \item \textbf{Nuclear energy:} debating whether nuclear power should be included in the national energy mix.  
    \item \textbf{Reddito di cittadinanza (Citizens' Income):} a state subsidy for individuals living in poverty, functioning as a conditional and non-individual guaranteed minimum income\footnote{\url{https://www.forbes.com/sites/annalisagirardi/2019/04/01/the-italian-citizens-income-reform-definition-and-adjustments/#3d3102b249db}}, designed to ensure a minimum standard of living and to promote employment integration. Debates around this policy revolved around three main stances: approval, reform, and abolishment.
\end{itemize}

Posts were presented to users through Y's recommender system option. 
In particular, we tested two configurations:

\begin{itemize}
    \item \textbf{ReverseChronoFollowersPopularity:} the default setting (henceforth \textit{Default}), where recent posts from followed users are shown, ranked by popularity, with some exposure to external users.  
    \item \textbf{ContentRecSys:} a random baseline (henceforth \textit{Random}), where posts are sampled randomly from all platform content.  
\end{itemize}

The recommender systems relies on the evolving network structure to distribute content among users, as the network evolved under the two extreme conditions we used as initialization strategies.

Finally, for each combination of parameters (model, network initialization strategy, and recommender system), simulations were run for \textbf{21 virtual days} with \textbf{100 agents}, and each condition was repeated \textbf{10 times} to ensure statistical robustness.

\subsubsection{Opinion modelling}
In addition to the simulation-driven opinion evolution, we are considering a well-grounded in literature approach: the Friedkin–Johnsen model~\cite{friedkin_1990} is a foundational framework in opinion dynamics that extends the classical DeGroot~\cite{degroot1974reaching} model by incorporating the concept of individual resistance to change. 
In this model, each agent updates their opinion as a weighted combination of their neighbors’ views and their own initial belief, with a susceptibility parameter determining the balance between external influence and internal stubbornness. 
We choose the Friedkin–Johnsen model because it generalizes the DeGroot framework while relaxing some of its restrictive assumptions, allowing opinions to stabilize without necessarily converging top consensus. Moreover, unlike bounded-confidence models such as Hegselmann–Krause~\cite{HegselmannRea02_OpinionDynamicsBounded}, it does not impose explicit clustering, making it well suited for scenarios where influence is continuous and heterogeneous rather than strictly segmented.
\subsubsection{Coalition distribution in the population}
\begin{figure}[!t]
    \centering
    \includegraphics[width=\linewidth]{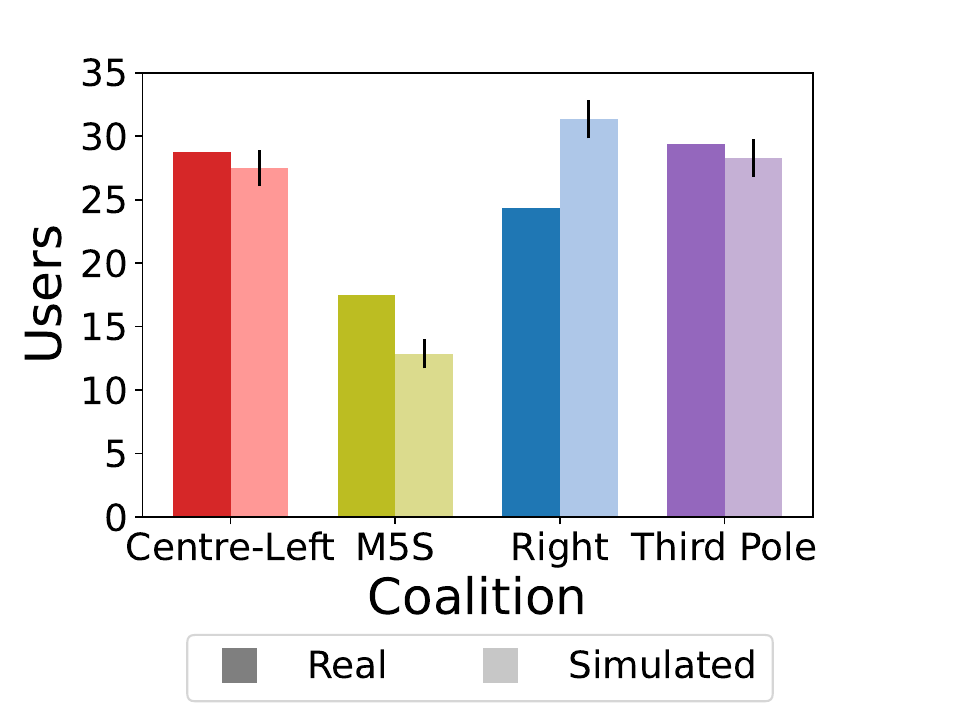}
    \caption{Percentages of users per political coalition in the real-world data and agents in the simulations.
    For the simulations, each bar represents the mean value across simulation runs with confidence $0.95$.}
    \label{fig:population}
\end{figure}

At initialization, users were assigned political leanings by sampling from real-world data, reproducing the coalition distribution observed in the source dataset.
As shown in Figure~\ref{fig:population}, the population is imbalanced: the Right coalition dominates, Centre-Left and Third Pole are of comparable size, and M5S (Movimento 5 Stelle) is smaller with low variability across simulations.
More information about the Italian political coalitions can be found in Appendix \ref{app:coalition_opinions}.
\subsubsection{Toxicity analysis}
Content toxicity was assessed using the Detoxify library~\cite{hanu2020detoxify}, which provides a continuous score between 0 and 1.
We analyzed toxicity in our simulations both in relation to interactions with in-group versus out-group users.

\subsection{In-group and out-group interactions between agents} \label{subsec:interactions}
We begin by examining patterns of inter-group interactions among agents, divided into in-group and out-group interactions; specifically, the frequency with which agents engage with others affiliated with the same political coalition versus those aligned with opposing coalitions. 
To assess the similarity between simulated and real-world behavior, we aggregate interactions from 10 independent runs for each parameter configuration (model, network initialization, and recommender system) into a single dataset. 
We then compute the proportion of in-group and out-group interactions for each coalition, applying the same procedure to the empirical interaction data used to initialize the simulations. 
In particular, we focus on replies. 
Finally, we evaluate correspondence by calculating Pearson correlations: four values capturing in-group alignment (one per coalition) and twelve values capturing out-group interactions (three per coalition).
For in-group interactions, we compute the correlations between the diagonals of the simulated and empirical interaction matrices. These values capture the degree to which the simulated model reproduces the proportion of interactions occurring among users supporting the same political coalition.

For out-group interactions, we instead focus on the off-diagonal elements of the interaction matrices, which represent interactions between users of different political leanings. In this case, the correlations quantify how well the simulated percentages of cross-coalition interactions align with those observed in real data.

\begin{figure}[!t]
    \centering
    \includegraphics[width=\linewidth]{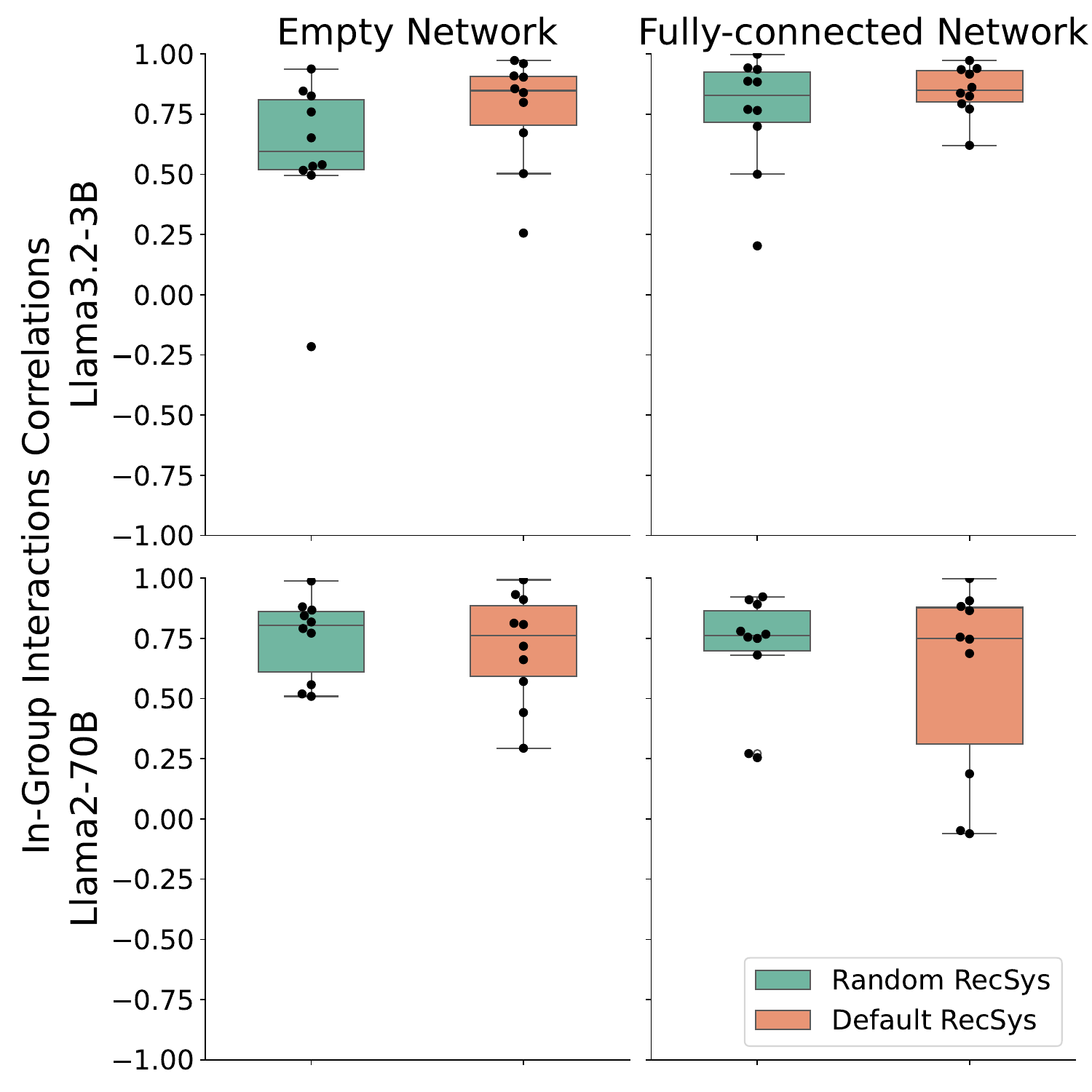}
    \caption{Distribution of the Pearson correlations of in-group interactions between the simulation and the real data, divided by model and network initialization strategy. Each dot is a simulation. Overall, approximately 91\% of the correlations are not significant ($p>0.05$). Similar results are obtained using Spearman $\rho$.}
    \label{fig:interactions_in}
\end{figure}

\begin{figure}[!t]
    \centering
    \includegraphics[width=\linewidth]{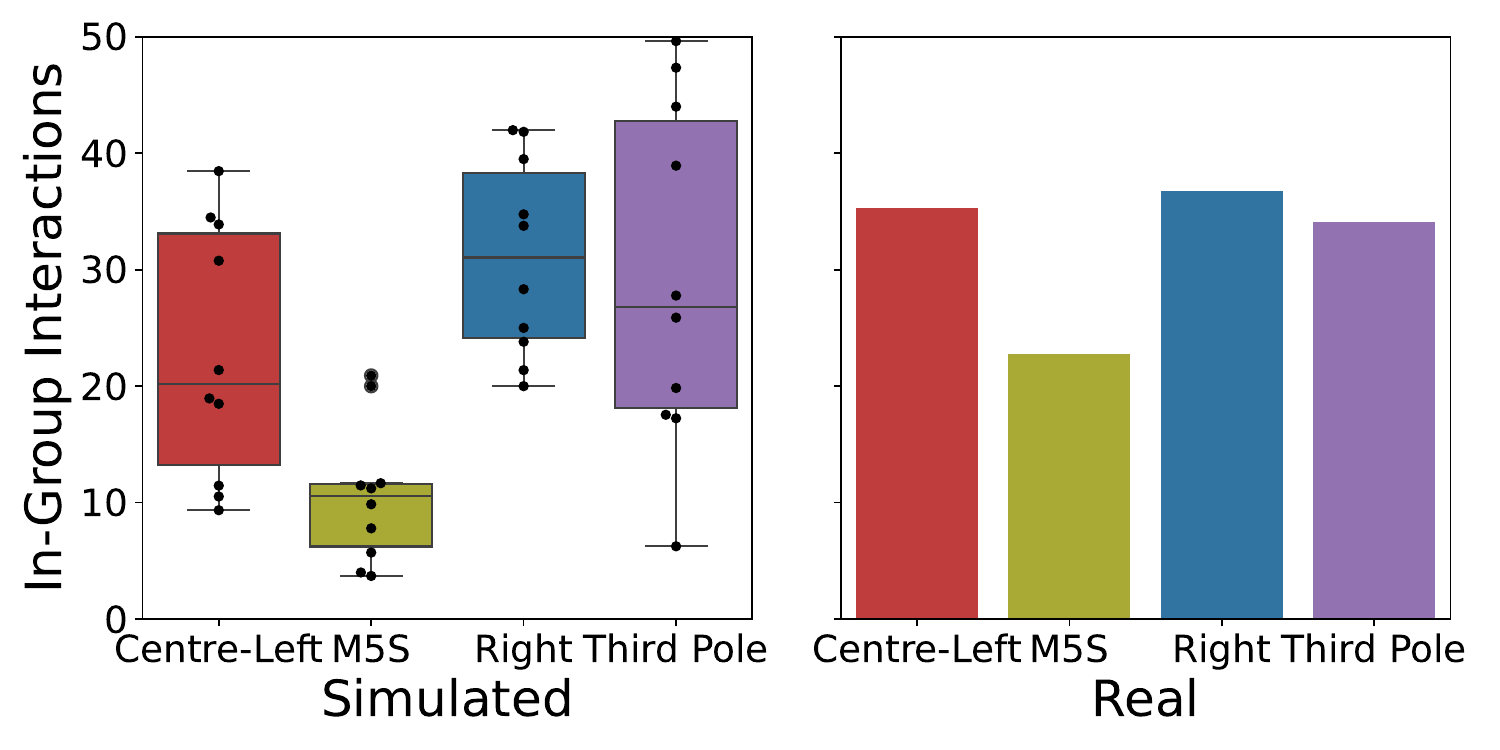}
    \caption{Example of comparison between simulated (10 runs across 1 configuration) and real data. On the left, the distribution of in-group interactions percentages for each coalition (using \artiname{}, empty initial network, and random recommender system). On the right, the percentages of in-group interactions for each coalition in the real-world dataset.}
    \label{fig:interactions_in_eg}
\end{figure}

Figure~\ref{fig:interactions_in} reports the distribution of correlations between simulated and real in-group interaction frequencies, disaggregated by model type and network initialization. 
Focusing first on the smaller model (top row), we observe that the recommender system exerts the strongest influence. 
In particular, the Default system substantially improves alignment with the real data, raising the median correlation to 0.85, an increase of almost 30 percentage points over the Random system (median = 0.6). 
While some runs with the Random system achieve similar levels of accuracy, its outcomes are far less stable, with correlations falling as low as –0.28 when simulations begin from an empty network. 
This gap narrows under the fully connected initialization, where the two recommenders yield similar median correlations; however, variance remains consistently higher in the Random condition, except for \llname{} in the Default system that exhibits outliers with negative correlation. Overall, however, the simulations tend to reproduce fairly well the extent to which users interact homophilously, at least on average.
To illustrate the analysis in more detail, Figure~\ref{fig:interactions_in_eg} presents a case study of a single experimental configuration (\artiname{}, empty initialization, Random recommender). 
Here, we compare the distribution of simulated in-group interactions for each coalition against the corresponding empirical proportions. 
The summary results in Figure~\ref{fig:interactions_in} are then derived by computing, for each run, the Pearson correlation between the four coalition-level values produced by the simulation and those observed in the real-world dataset.

\begin{figure}[!t]
    \centering
    \includegraphics[width=\linewidth]{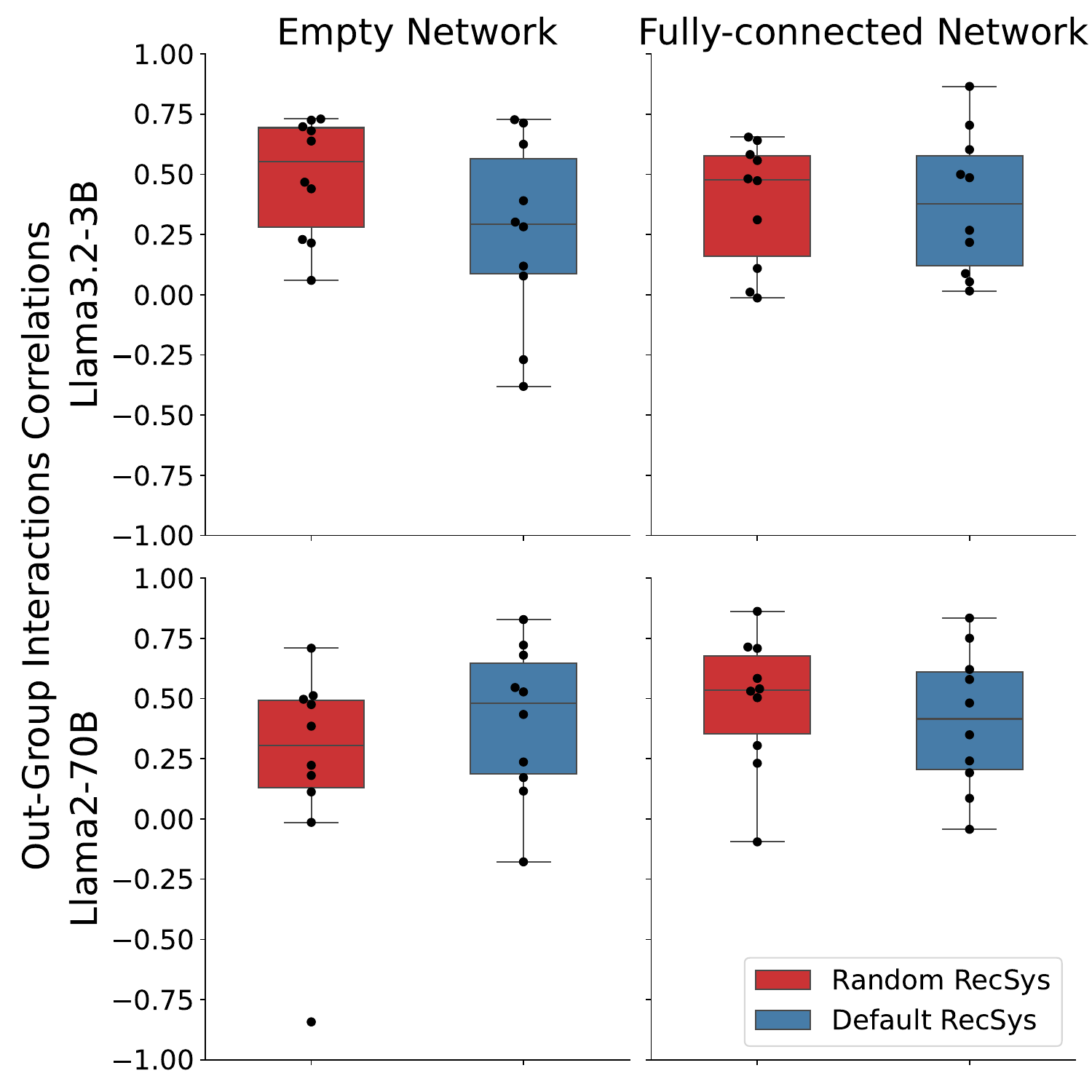}
    \caption{Distribution of the Pearson correlations of out-group interactions between the simulation and the real data, divided by model and network initialization strategy. Each dot is a simulation. Overall, approximately 66\% of the correlations are not significant ($p>0.05$). Similar results are obtained using Spearman $\rho$.}
    \label{fig:interactions_out}
\end{figure}

Figure~\ref{fig:interactions_out} reports, for each configuration of model and network initialization, the distribution of correlations between simulated and real out-group interactions. Compared to in-group interactions, the degree of fidelity is noticeably lower: median correlations rarely exceed 0.6, indicating that cross-coalition dynamics are more difficult to replicate.  

\noindent Differences across models, initialization strategies, and recommender systems are relatively small, with no configuration standing out as consistently superior. Nevertheless, some individual runs yield extreme values, with correlations ranging from strongly negative to as high as 0.8, underscoring the presence of outliers and the instability of certain setups.  

Overall, these results suggest that while homophilous interactions tend to be faithfully reproduced, capturing out-group dynamics remains more challenging, with only moderate similarity to real-world data and substantial variability across simulation runs.

To clarify how out-group interaction patterns are compared between simulations and real data, we briefly outline the procedure here and provide a concrete example in Appendix~\ref{app:plots} (Figure~\ref{fig:interactions_out_eg}). For a fixed configuration (\artiname{}, empty initial network, Random recommender), we construct an inter-coalition interaction matrix from the simulations and an analogous matrix from the empirical dataset, both normalized to represent the proportion of interactions flowing from each coalition to every other coalition.
For each simulation run, we extract the twelve off-diagonal entries of the simulated matrix, corresponding to out-group interactions, and directly compare them with the matching twelve empirical values. This produces a single correlation score per run, quantifying how closely the simulated interaction structure aligns with the observed one. The summary correlations obtained through this procedure, aggregated across runs and configurations, are reported in Figure~\ref{fig:interactions_out}.

Overall, these experiments highlight the nuanced roles of recommender systems, models, and network initialization strategies in shaping the fidelity of simulations. 
Interestingly, altering these parameters does not produce large shifts in overall similarity. 
In-group interactions remain consistently the most faithfully reproduced, with high median correlations across configurations. 
By contrast, out-group interactions are considerably harder to replicate, with median scores ranging only from 0.17 to 0.4.

\subsection{In- and out-group toxic behaviour} \label{subsec:toxicity}
We now explore the toxicity of the comments in the simulations.
Our goal is to evaluate whether the simulated environments produce similar patterns of hostile communication observed in the real-world data.
To this end, we measure toxicity by computing the toxicity score of all replies and extracting the $95^\mathsf{th}$ percentile for each parameter configuration. 
We then distinguish between toxicity directed at members of the same coalition (in-group toxicity) and toxicity directed at members of other coalitions (out-group toxicity).  

\begin{figure}[!t]
    \centering
    \includegraphics[width=\linewidth]{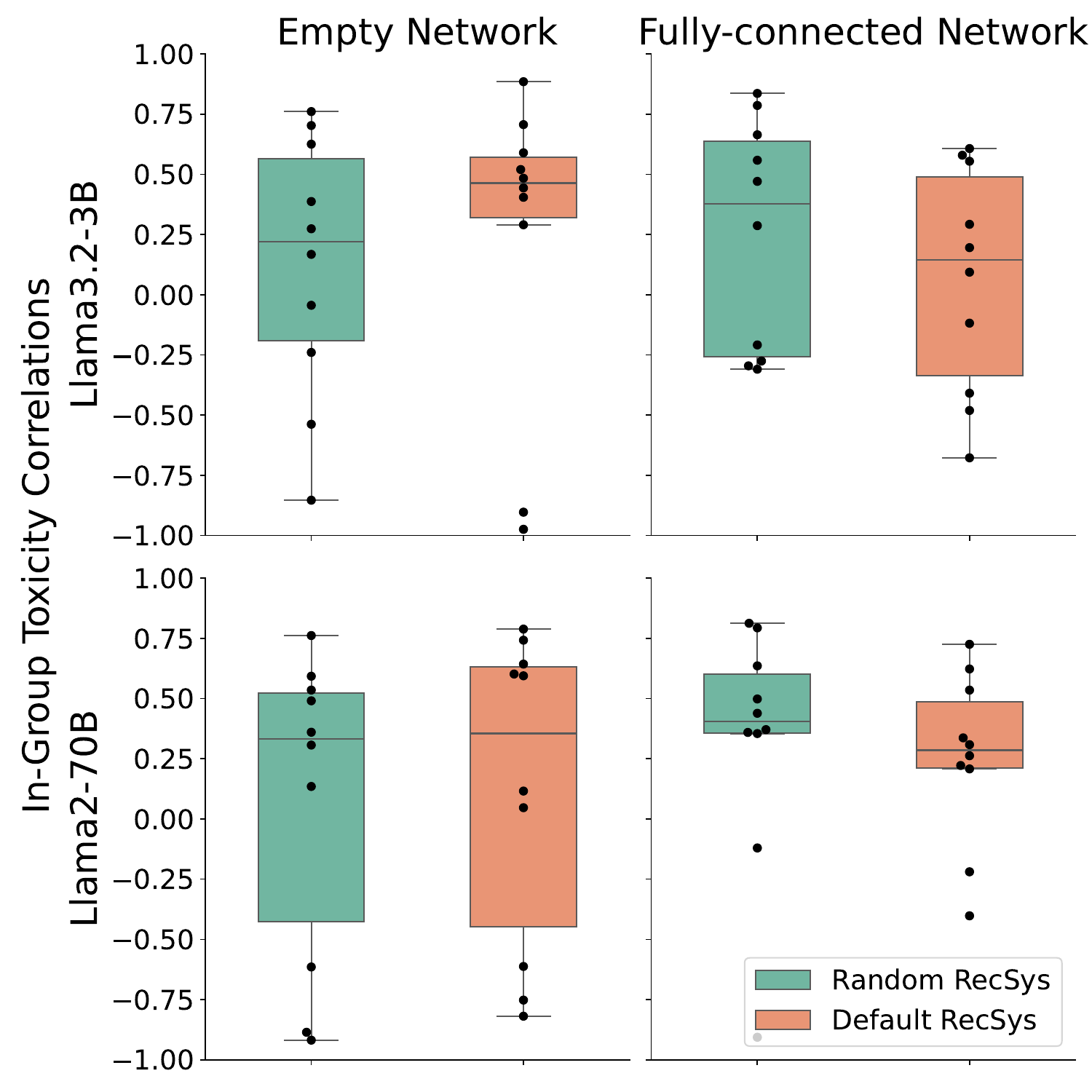}
    \caption{
    Distribution of the Pearson correlations of in-group toxicity between the simulation and the real data, divided by model and network initialization strategy. Each dot is a simulation. Overall, approximately 98\% of the correlations are not significant ($p>0.05$). Similar results are obtained using Spearman $\rho$.}
    \label{fig:toxicity_in}
\end{figure}

To assess how faithfully the simulations capture these patterns, we adopt the same strategy used in the interaction analyses: we correlate the simulated toxicity values with those observed in the empirical dataset used to initialize the runs.
Figure~\ref{fig:toxicity_in} reports the resulting correlations for in-group toxicity, divided by recommender system and network initialization strategy.  

Across most configurations, we observe high variability in the simulations. 
The only clear exception is \llname{} with a fully connected initialization, which yields a compact distribution of correlations with relatively high values under both recommendation strategies. 
In other settings, the most stable configuration is the combination of the Default recommender with an empty starting network, which achieves the highest median correlation overall.

\begin{figure}[!t]
    \centering
    \includegraphics[width=\linewidth]{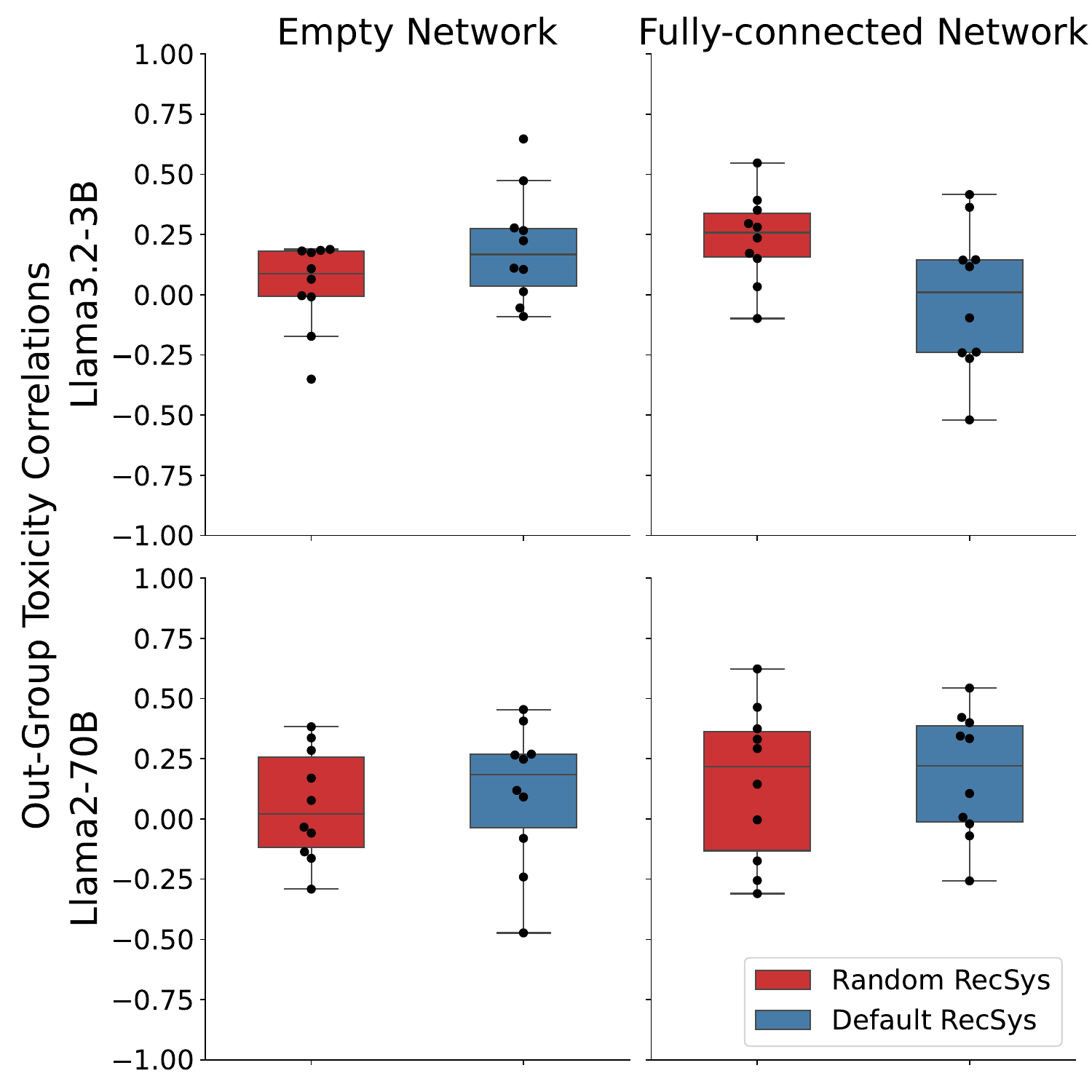}
    \caption{
   Distribution of the Pearson correlations of out-group toxicity between the simulation and the real data, divided by model and network initialization strategy. Each dot is a simulation. Overall, approximately 97\% of the correlations are not significant ($p>0.05$). Similar results are obtained using Spearman $\rho$.
    }
    \label{fig:toxicity_out}
\end{figure}

Turning to out-group toxicity (Figure \ref{fig:toxicity_out}), we find less variability across runs and also a similar trend with respect to network initialization compared to in-group analysis. 
With an empty network initialization, the Default algorithm produces high correlations with the real data, whereas under the fully connected initialization it performs worse.
The choice of model appears to have little influence on the outcome, as median correlation values remain largely comparable across the two uncensored models.  
Taken together, these findings suggest that simulations struggle to replicate patterns of toxic behavior among online users, despite being based on uncensored models. As in the case of interaction analyses, illustrative examples of preliminary investigations are presented in Figure~\ref{fig:toxicity_in_eg} and in the Appendix~\ref{app:plots} (Figure~\ref{fig:toxicity_out_eg}).
We see that, while the real distribution of in-group toxicity is evenly present across all political coalitions, the models fail to reproduce this phenomenon, with the most significant cases being M5S coalition and Center-Left.
Similarly, the same pattern emerges for overall out-group toxicity, where some trends are correctly replicated, while others (e.g., Third Pole vs. M5S) are greatly over estimated or under estimated (the visual presentation of this is visible in Appendix~\ref{app:plots}, Figure~\ref{fig:toxicity_out_eg}).

\begin{figure}[!t]
    \centering
    \includegraphics[width=\linewidth]{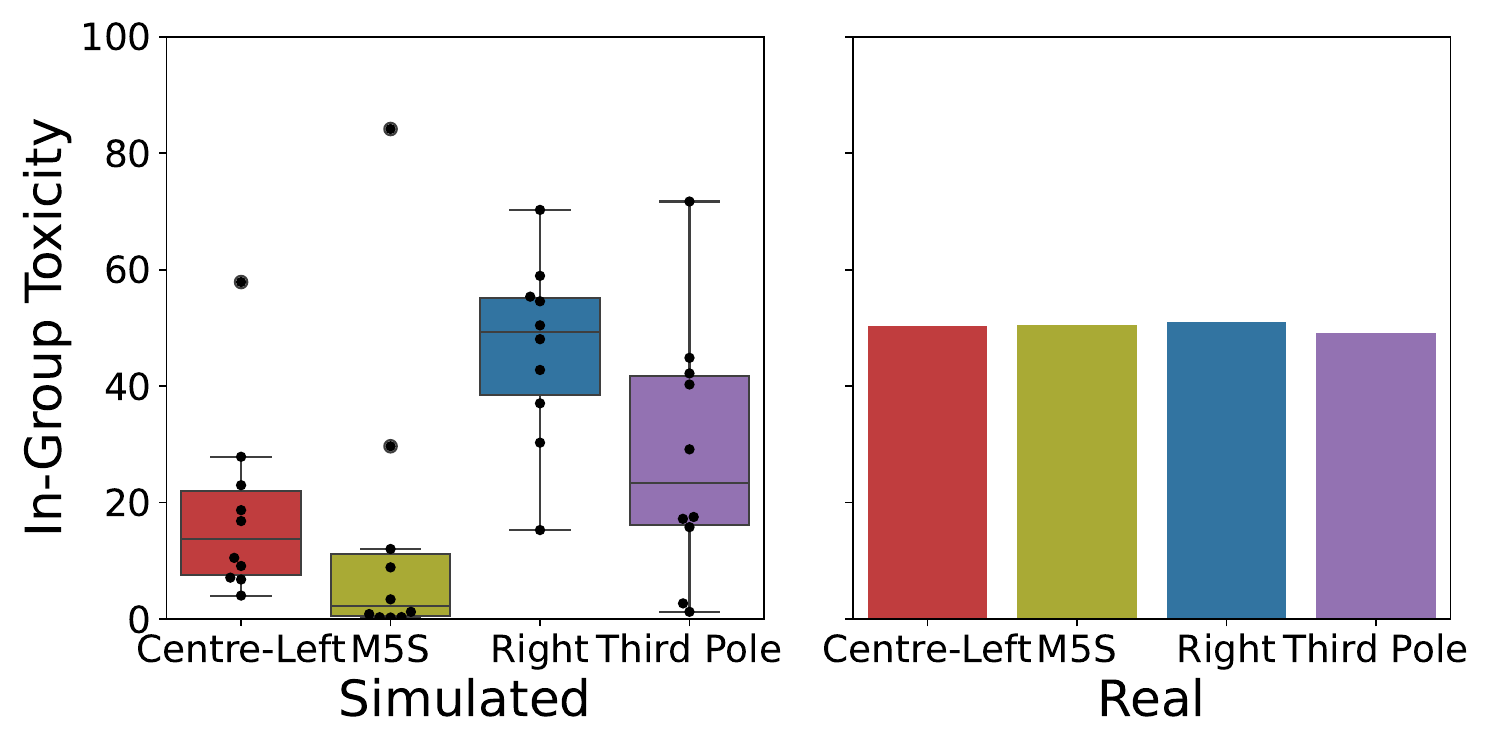}
    \caption{Example of comparison between simulated (10 runs across 1 configuration) and real data. On the left, the distribution of in-group toxicity percentages for each coalition (using \artiname{}, empty initial network, and random recommender system). On the right, the percentages of in-group toxicity for each coalition in the real-world dataset.}
    \label{fig:toxicity_in_eg}
\end{figure}

\subsection{Opinion dynamics} \label{subsec:opinion_dynamics}
\begin{figure}[!t]
    \centering
    \includegraphics[width=1\linewidth]{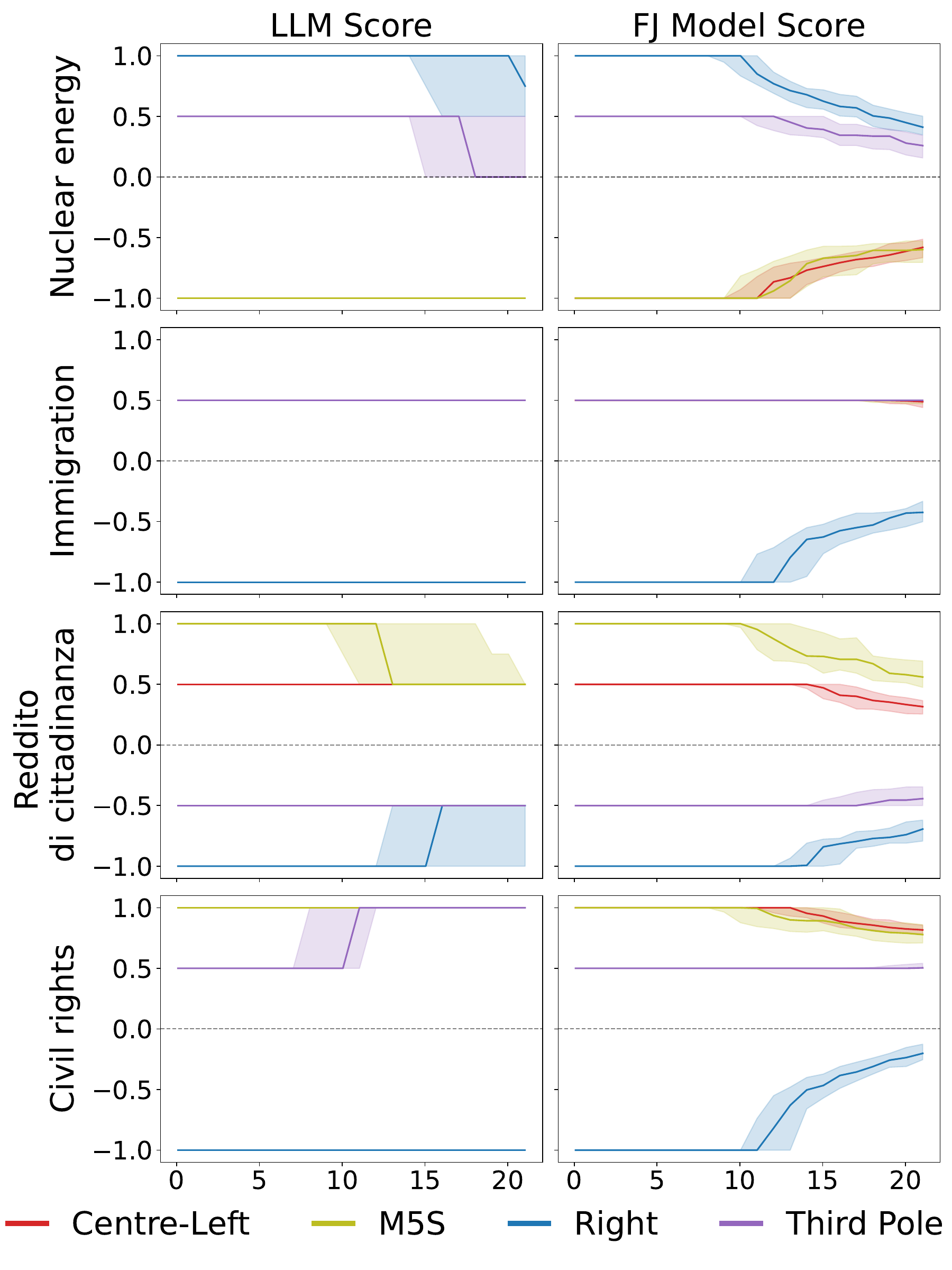}
    \caption{Example of evolution of opinion for each topic, comparing LLM-assigned score (\textit{LLM Score}, left column) and the one assigned by the traditional Friedkin–Johnsen model (\textit{FJ Model Score}, right column).
    Each line represents the median opinion on each day for users belonging to a certain coalition, along with a 95\% confidence interval.
    The data is aggregated over all simulation runs of a single experimental setup (\artiname{}, empty network, and random recommender system, in this case). Additional examples are provided in the Appendix.}
    \label{fig:opinion_evolution}
\end{figure}

We now examine how opinions evolve over time within the simulations, an important dimension for assessing whether LLM-based agents can approximate established models of opinion dynamics. In particular, we compare opinion trajectories across topics and coalitions against the predictions of the classical Friedkin--Johnsen (FJ) model, a widely used benchmark for modeling opinion change.  
Figure~\ref{fig:opinion_evolution} illustrates this comparison for a single setup, contrasting opinion scores assigned by LLMs (left column) with those generated by the FJ model (right column). Overall, both approaches display broadly coherent trends: opinions evolve with comparable trajectories and frequently converge toward similar mean values across coalitions.  

The most salient difference lies in the pace and smoothness of change. For example, in the case of \texttt{Nuclear Energy}, both the \textit{Third Pole} and the \textit{Right} coalitions converge toward neutrality, yet the FJ model shows a gradual adjustment whereas the LLM-based simulation produces a sharper shift, suggesting a limitation in capturing incremental opinion change. A similar pattern emerges in \texttt{Reddito di Cittadinanza}, where \textit{Right} and \textit{M5S} converge toward neutral positions with noticeably different slopes. 
In this case, the FJ model also captures a ``neutralization'' effect for the \textit{Centre-Left} and \textit{Third Pole}, which the LLM-based approach fails to reproduce. 
A likely explanation is the difficulty LLMs face in handling fine-grained scoring: their outputs often resemble step functions rather than continuous curves.  

The largest divergence appears in the case of \texttt{Civil Rights}. In the LLM-based simulation, the \textit{Third Pole} shifts toward full approval, while all other coalitions remain static. 
By contrast, the FJ model predicts gradual changes for multiple groups: \textit{Centre-Left}, \textit{M5S}, and especially the \textit{Right} all move toward neutrality, each along distinct trajectories.  

Taken together, these findings suggest that LLM-based simulations can reproduce opinion change at the population level, as their aggregate behavior is often comparable to that of established models. 
However, they display lower sensitivity and a tendency toward abrupt adjustments rather than incremental change. 
Notably, coalitions that start from identical positions exhibit perfectly overlapping trends, indicating that initial opinions dominate the dynamics more strongly in the LLM-based setting.  

\begin{figure}[!t]
    \centering
    \includegraphics[width=\linewidth]{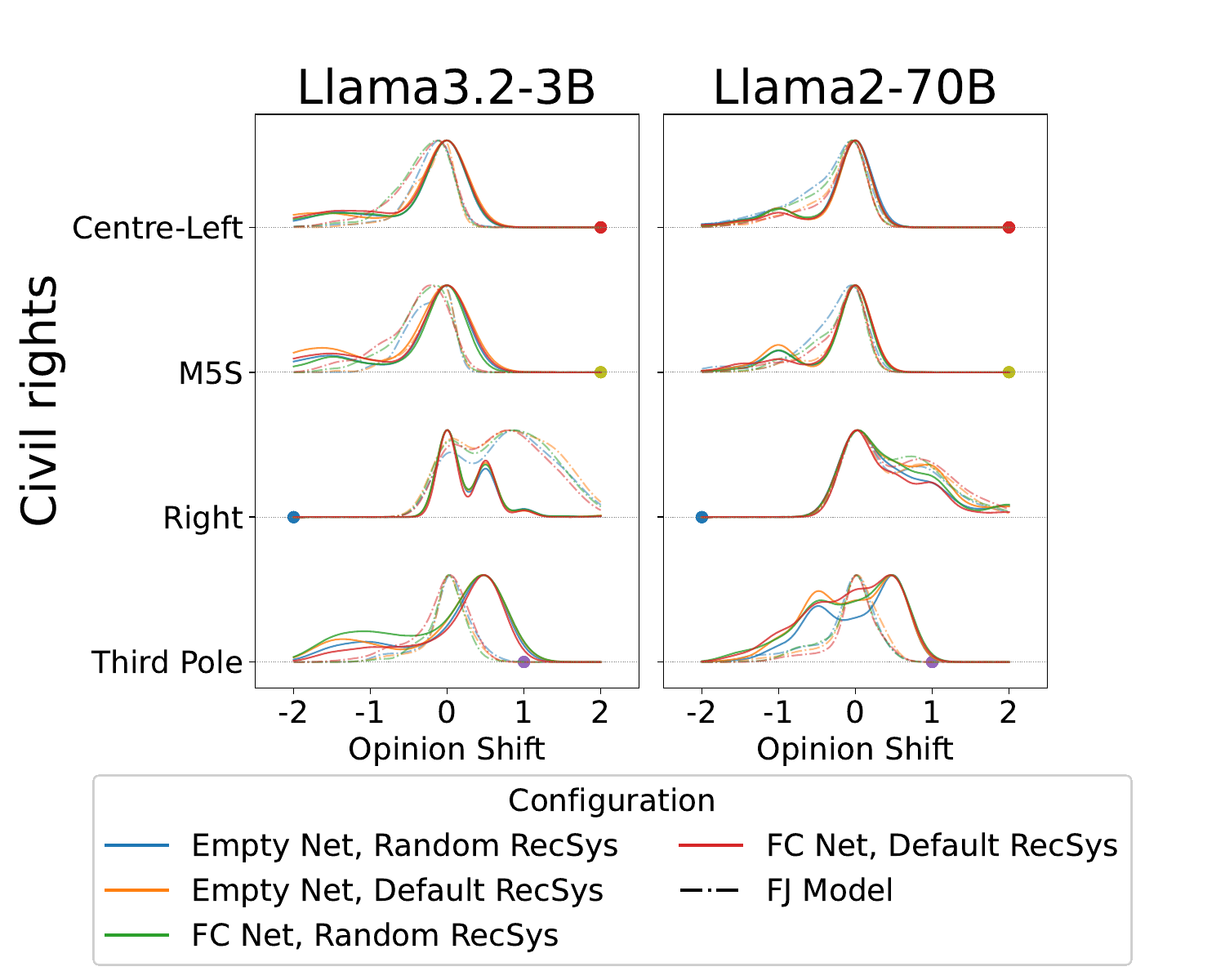}
    \caption{Example of opinion shifts for each coalition for different configurations of model, network initialization, and recommender system. For each configuration, the corresponding simulation using the Friedkin–Johnsen mathematical model is reported (dashed lines).
    Initial opinions are highlighted for each party and for each topic. $2.0$ means \texttt{Strongly Supportive}, whereas $-2.0$ means \texttt{Strongly Opposed}.
    Additional plots for topics different from \texttt{Civil Rights} are provided in the Appendix.}
    \label{fig:opinion_shifts}
\end{figure}

Finally, Figure~\ref{fig:opinion_shifts} reports the relative opinion shifts for the \texttt{Civil Rights} topic across coalitions. 
Here again, the LLM-based simulations track the broad tendencies of the FJ model, but they fail to capture all nuances. 
In particular, the FJ model predicts a substantial shift by the \textit{Right} coalition, which is not reproduced by the LLMs. 
This discrepancy highlights the current limitations of LLMs in capturing the subtle and heterogeneous dynamics of opinion change.  

As shown in Appendix~\ref{app:plots}, these dynamics are consistent across topics: opinions progressively converge toward neutral values, reflecting a decline in polarization over time. 
Whether this trend persists or stabilizes with longer simulations remains an open question. 
Importantly, the same pattern is observed across models, network structures, and recommender systems, underscoring the robustness of the result.  

\section{Discussion and Conclusion}\label{sec:conclusion}

LLMs have emerged as a promising tool for simulating agents in virtual environments. The main goal of this work was to provide an exploratory analysis of the behavior of LLM-based agents in the context of online social media platforms. To this end, we extended the \textit{Y} simulator by integrating mechanisms for opinion evolution and introducing a realistic user initialization procedure grounded in empirical data from the 2022 Italian political context. This framework enabled us to systematically test how different generative models, network initialization strategies, and recommender systems affect the fidelity of the simulations when compared with real-world observations.  

To explore the behavior of simulated agents from multiple perspectives, we conducted analyses at three levels: interaction patterns, opinion dynamics, and the toxicity of generated content. Overall, our findings indicate that LLMs represent a promising approach for simulating user behavior online. Agents successfully interacted, formed connections, and produced content with varying levels of toxicity, though their communication style was systematically biased toward neutrality.

At the same time, several limitations should be acknowledged. The 21 simulated days were sufficient for a network structure to begin forming, but too short to capture longer-term or emergent dynamics. For instance, actions such as \textit{unfollow} were almost absent, and the full effects of recommendation algorithms may not have materialized given the relatively unstructured networks in the early phases of simulation.
Future work should address this by implementing more faithful initialization techniques, such as Stochastic Block Models, to generate networks with a realistic density and predefined community structure. 
This will enable a more robust and nuanced study of emergent dynamics within established social environments.
With respect to opinion evolution, LLM-assigned scores were broadly consistent with those produced by traditional models, both showing a tendency to converge toward neutral positions. However, the relatively short duration leaves open whether these trends would stabilize, polarize, or diverge over longer time horizons. Our choice of simulation length was ultimately constrained by available computational resources.  

Although simulated interaction and toxicity patterns were not significantly correlated with empirical data, this result should be interpreted in light of the exploratory and short-horizon nature of the simulations rather than as a failure of the modeling approach. The lack of significant correlations is in contrast with previous findings, particularly~\citet{gao2023s3socialnetworksimulationlarge}, showing that it is crucial for both traditional opinion dynamics models and LLM-based agents to have longer time horizons and more structured networks to produce stable, macro-level alignment with empirical observations. From this perspective, our results suggest that current LLM-based simulations can be considered still in their early stage of development and far from being suited for direct quantitative prediction of real-world interaction patterns.

Nevertheless, despite the limited temporal scope of our simulations, our results are consistent with those reported by~\citet{chuang2024simulatingopiniondynamicsnetworks} with respect to opinion evolution. In particular, agents exhibit a clear tendency to converge toward neutral positions. This behavior aligns with previous findings showing that, in the absence of explicitly modeled cognitive biases, opinion dynamics tend to stabilize rather than polarize.

Future developments should focus on enriching agent personalization and studying the robustness of these systems. Incorporating elements such as emotional reasoning, susceptibility to influence, or varying levels of trust in consumed information could enable more realistic opinion and interaction dynamics.
We also relied on a fixed prompt structure. Considering the critical role that prompt sensitivity plays in LLM–based experiments, future work should perform ablation studies to assess how sensitive agent behavior is to variations in prompt phrasing. 
Another important extension is the inclusion of external shocks—such as social crises~\cite{di2022vaccineu}, scandals, or major public statements—to evaluate how agents respond to events that typically shape online discourse. 
It could also be of interest to see the impact of simulated harmful content, such as conspiracy theories~\cite{corso2025early}, misinformation, disinformation~\cite{nogara2026longitudinal}, or political advertisement~\cite{pierri2023political}.
More systematic comparisons with empirical data are also necessary to better evaluate the realism of emergent behaviors. Finally, expanding beyond the Italian political context would allow us to assess the generalizability of the approach and test whether the observed dynamics hold across different sociopolitical settings.  
In conclusion, integrating LLMs as agents in social simulations represents an important step toward more realistic modeling of online environments, particularly with respect to language, interactions, and content generation. However, replicating more heterogeneous phenomena, such as the spread of misinformation, will require further advances in behavioral modeling. Our study contributes to the growing exploration of LLM-driven simulations, providing a foundation for investigating complex social dynamics under controlled conditions.

\section*{Ethical Statement}\label{sec:ethical}

The deployment of Large Language Models as AI social agents raises numerous ethical considerations that are currently the subject of intense scrutiny by the interdisciplinary research community. The extraordinary capabilities of these models to generate text have led several scientists to envision alarming scenarios in which the seamless integration of AI agents into the online social discourse may facilitate the dissemination of harmful content, the spread of misinformation, and the propagation of `semantic garbage', ultimately damaging our societies~\cite{FloridiLea20_NatureScopesLimitsConseqsGPT3, WeidingerLea22_TaxonomyRisksPosed, HendrycksDea23p_OverviewCatastrophicAI}. 
As a result, any research exploring the characteristics of LLMs as social agents could, directly or indirectly, contribute knowledge that might be exploited to implement and deploy LLM-based technologies for malicious purposes. While recognizing this risk, we also believe that conducting research on LLM-based agents is essential to assess potential risks and to guide efforts aimed at developing strategies to mitigate them. Our study contributes positively to deepen our understanding of how LLMs react to social stimuli.

Even when deploying LLM-based agents for ethical purposes, trade-offs between the obtained benefit and the high level of power consumption required to run them should be carefully considered~\cite{BenderEea21_LLMsAsStochasticParrots}.

\section{Acknowledgments}
We acknowledge ISCRA for awarding this project access to the LEONARDO supercomputer, owned by the EuroHPC Joint Undertaking, hosted by CINECA (Italy).
Francesco Pierri is partially supported by PNRR-PE-AI FAIR project funded by the NextGeneration EU program.

\bibliography{bibliography}

@misc{rossetti2024ysocialllmpoweredsocial,
    title={Y Social: an LLM-powered Social Media Digital Twin}, 
    author={Giulio Rossetti and Massimo Stella and Rémy Cazabet and Katherine Abramski and Erica Cau and Salvatore Citraro and Andrea Failla and Riccardo Improta and Virginia Morini and Valentina Pansanella},
    year={2024},
    eprint={2408.00818},
    archivePrefix={arXiv},
    primaryClass={cs.AI},
    url={https://arxiv.org/abs/2408.00818}, 
}

@misc{gao2023s3socialnetworksimulationlarge,
    title={S3: Social-network Simulation System with Large Language Model-Empowered Agents}, 
    author={Chen Gao and Xiaochong Lan and Zhihong Lu and Jinzhu Mao and Jinghua Piao and Huandong Wang and Depeng Jin and Yong Li},
    year={2023},
    eprint={2307.14984},
    archivePrefix={arXiv},
    primaryClass={cs.SI},
    url={https://arxiv.org/abs/2307.14984}, 
}

@misc{törnberg2023evaluate,
    title={Simulating Social Media Using Large Language Models to Evaluate Alternative News Feed Algorithms}, 
    author={Petter Törnberg and Diliara Valeeva and Justus Uitermark and Christopher Bail},
    year={2023},
    eprint={2310.05984},
    archivePrefix={arXiv},
    primaryClass={cs.SI},
    url={https://arxiv.org/abs/2310.05984}, 
}

@article{friedkin_1990,
    author = {Friedkin, Noah and Johnsen, Eugene},
    year = {1990},
    month = {01},
    pages = {193-206},
    title = {Social Influence and Opinions},
    volume = {15},
    journal = {Journal of Mathematical Sociology - J MATH SOCIOL},
    doi = {10.1080/0022250X.1990.9990069}
}

@inproceedings{Liu_2018,
   title={Discrete-Time Polar Opinion Dynamics with Heterogeneous Individuals},
   url={http://dx.doi.org/10.1109/CDC.2018.8619071},
   DOI={10.1109/cdc.2018.8619071},
   booktitle={2018 IEEE Conference on Decision and Control (CDC)},
   publisher={IEEE},
   author={Liu, Ji and Ye, Mengbin and Anderson, Brian D.O. and Basar, Tamer and Nedic, Angelia},
   year={2018},
   month=dec, pages={1694–1699} 
}

@inproceedings{Ye2018Opinion,
    author    = {Mengbin Ye and Ji Liu and Brian D. O. Anderson},
    title     = {Opinion Dynamics with State-Dependent Susceptibility to Influence},
    booktitle = {Proceedings of the 23rd International Symposium on Mathematical Theory of Networks and Systems (MTNS)},
    year      = {2018},
    url       = {https://mtns2018.hkust.edu.hk/media/files/0044.pdf}
}

@misc{cau2025languagedrivenopiniondynamicsagentbased,
    title={Language-Driven Opinion Dynamics in Agent-Based Simulations with LLMs}, 
    author={Erica Cau and Valentina Pansanella and Dino Pedreschi and Giulio Rossetti},
    year={2025},
    eprint={2502.19098},
    archivePrefix={arXiv},
    primaryClass={cs.SI},
    url={https://arxiv.org/abs/2502.19098}, 
}

@inproceedings{chuang2024simulatingopiniondynamicsnetworks, 
    address={Mexico City, Mexico}, 
    title={Simulating Opinion Dynamics with Networks of LLM-based Agents}, 
    url={https://aclanthology.org/2024.findings-naacl.211/}, 
    DOI={10.18653/v1/2024.findings-naacl.211}, 
    booktitle={Findings of the Association for Computational Linguistics: NAACL 2024}, 
    publisher={Association for Computational Linguistics}, 
    author={Chuang, Yun-Shiuan and Goyal, Agam and Harlalka, Nikunj and Suresh, Siddharth and Hawkins, Robert and Yang, Sijia and Shah, Dhavan and Hu, Junjie and Rogers, Timothy}, 
    editor={Duh, Kevin and Gomez, Helena and Bethard, Steven}, 
    year={2024}, 
    month=jun, 
    pages={3326–3346}
}

@inproceedings{Liu_2024, series={IJCAI-2024},
    title={From Skepticism to Acceptance: Simulating the Attitude Dynamics Toward Fake News},
    url={http://dx.doi.org/10.24963/ijcai.2024/873},
    DOI={10.24963/ijcai.2024/873},
    booktitle={Proceedings of the Thirty-ThirdInternational Joint Conference on Artificial Intelligence},
    publisher={International Joint Conferences on Artificial Intelligence Organization},
    author={Liu, Yuhan and Chen, Xiuying and Zhang, Xiaoqing and Gao, Xing and Zhang, Ji and Yan, Rui},
    year={2024},
    month=aug, pages={7886–7894},
    collection={IJCAI-2024} 
}

@misc{piao2025emergencehumanlikepolarizationlarge,
    title={Emergence of human-like polarization among large language model agents}, 
    author={Jinghua Piao and Zhihong Lu and Chen Gao and Fengli Xu and Qinghua Hu and Fernando P. Santos and Yong Li and James Evans},
    year={2025},
    eprint={2501.05171},
    archivePrefix={arXiv},
    primaryClass={cs.SI},
    url={https://arxiv.org/abs/2501.05171}, 
}

@article{hu2025simulatingrumorspreadingsocial,
  title={Simulating rumor spreading in social networks using llm agents},
  author={Hu, Tianrui and Liakopoulos, Dimitrios and Wei, Xiwen and Marculescu, Radu and Yadwadkar, Neeraja J},
  journal={arXiv preprint arXiv:2502.01450},
  year={2025}
}

@article{McCrae1992,
    author    = {Robert R. McCrae and Oliver P. John},
    title     = {An Introduction to the Five-Factor Model and Its Applications},
    journal   = {Journal of Personality},
    year      = {1992},
    volume    = {60},
    number    = {2},
    pages     = {175--215},
    month     = jun,
    doi       = {10.1111/j.1467-6494.1992.tb00970.x},
    pmid      = {1635039}
}

@article{macy2002abm,
    author       = {Michael W. Macy and Robb Willer},
    title        = {From Factors to Actors: Computational Sociology and Agent-Based Modeling},
    journal      = {Annual Review of Sociology},
    volume       = {28},
    pages        = {143--166},
    year         = {2002},
    doi          = {10.1146/annurev.soc.28.110601.141117},
    url          = {https://doi.org/10.1146/annurev.soc.28.110601.141117}
}

@article{squazzoni2014socialsimulation,
    author = {Flaminio Squazzoni and Wander Jager and Bruce Edmonds},
    title ={Social Simulation in the Social Sciences: A Brief Overview},
    journal = {Social Science Computer Review},
    volume = {32},
    number = {3},
    pages = {279-294},
    year = {2014},
    doi = {10.1177/0894439313512975},
    URL = {https://doi.org/10.1177/0894439313512975},
    eprint = {https://doi.org/10.1177/0894439313512975}
}

@article{conte2014agent,
    author  = {Rosaria Conte and Mario Paolucci},
    title   = {On Agent-Based Modeling and Computational Social Science},
    journal = {Frontiers in Psychology},
    volume  = {5},
    year    = {2014},
    doi     = {10.3389/fpsyg.2014.00668},
    url     = {https://www.frontiersin.org/articles/10.3389/fpsyg.2014.00668},
    issn    = {1664-1078}
}

@inproceedings{park2023genagents,
  title={Generative agents: Interactive simulacra of human behavior},
  author={Park, Joon Sung and O'Brien, Joseph and Cai, Carrie Jun and Morris, Meredith Ringel and Liang, Percy and Bernstein, Michael S},
  booktitle={Proceedings of the 36th annual acm symposium on user interface software and technology},
  pages={1--22},
  year={2023}
}

@article{Degroot1974,
  author    = {Morris H. DeGroot},
  title     = {Reaching a Consensus},
  journal   = {Journal of the American Statistical Association},
  volume    = {69},
  number    = {345},
  pages     = {118--121},
  year      = {1974},
  doi       = {10.1080/01621459.1974.10480137},
  url       = {https://www.tandfonline.com/doi/abs/10.1080/01621459.1974.10480137},
  eprint    = {https://www.tandfonline.com/doi/pdf/10.1080/01621459.1974.10480137}
}

@misc{statista2024twitter,
    author       = {{Statista Research Department}},
    title        = {Distribution of users on Twitter worldwide as of January 2024, by age group and gender},
    year         = {2024},
    url          = {https://www.statista.com/statistics/1498204/distribution-of-users-on-twitter-worldwide-age-and-gender/},
    note         = {Accessed on June 7, 2025}
}

@inproceedings{pierri2023ita,
  title={Ita-election-2022: A multi-platform dataset of social media conversations around the 2022 italian general election},
  author={Pierri, Francesco and Liu, Geng and Ceri, Stefano},
  booktitle={Proceedings of the 32nd ACM International Conference on Information and Knowledge Management},
  pages={5386--5390},
  year={2023}
}

@article{barrick1991bigfive,
    author    = {Barrick, Murray R. and Mount, Michael K.},
    title     = {The Big Five personality dimensions and job performance: A meta-analysis},
    journal   = {Personnel Psychology},
    volume    = {44},
    number    = {1},
    pages     = {1--26},
    year      = {1991},
    doi       = {10.1111/j.1744-6570.1991.tb00688.x},
    url       = {https://doi.org/10.1111/j.1744-6570.1991.tb00688.x}
}

@article{pierri2024drivers,
    author    = {Francesco Pierri},
    title     = {Drivers of Hate Speech in Political Conversations on Twitter: The Case of the 2022 Italian General Election},
    journal   = {EPJ Data Science},
    volume    = {13},
    number    = {1},
    pages     = {63},
    year      = {2024},
    doi       = {10.1140/epjds/s13688-024-00501-1},
    url       = {https://doi.org/10.1140/epjds/s13688-024-00501-1}
}

@article{failla2024,
    title={“I’m in the Bluesky Tonight”: Insights from a year worth of social data},
    volume={19},
    ISSN={1932-6203},
    url={http://dx.doi.org/10.1371/journal.pone.0310330},
    DOI={10.1371/journal.pone.0310330},
    number={11},
    journal={PLOS ONE},
    publisher={Public Library of Science (PLoS)},
    author={Failla, Andrea and Rossetti, Giulio},
    editor={Saracco, Fabio},
    year={2024},
    month=nov, pages={e0310330} 
}

@article{bakshy2015,
    author = {Bakshy, Eytan and Messing, Solomon and Adamic, Lada},
    year = {2015},
    month = {05},
    pages = {},
    title = {Political science. Exposure to ideologically diverse news and opinion on Facebook},
    volume = {348},
    journal = {Science (New York, N.Y.)},
    doi = {10.1126/science.aaa1160}
}

@article{vosoughi2018spread,
  author  = {Soroush Vosoughi and Deb Roy and Sinan Aral},
  title   = {The spread of true and false news online},
  journal = {Science},
  volume  = {359},
  number  = {6380},
  pages   = {1146--1151},
  year    = {2018},
  doi     = {10.1126/science.aap9559},
  url     = {https://www.science.org/doi/abs/10.1126/science.aap9559},
  eprint  = {https://www.science.org/doi/pdf/10.1126/science.aap9559}
}

@misc{hanu2020detoxify,
  author       = {Daniel Hanu},
  title        = {Detoxify},
  year         = {2020},
  howpublished = {\url{https://github.com/unitaryai/detoxify}},
  note         = {Accessed on June 24, 2025}
}

@article{lazer2009computational,
    author = {David Lazer  and Alex Pentland  and Lada Adamic  and Sinan Aral  and Albert-László Barabási  and Devon Brewer  and Nicholas Christakis  and Noshir Contractor  and James Fowler  and Myron Gutmann  and Tony Jebara  and Gary King  and Michael Macy  and Deb Roy  and Marshall Van Alstyne },
    title = {Computational Social Science},
    journal = {Science},
    volume = {323},
    number = {5915},
    pages = {721-723},
    year = {2009},
    doi = {10.1126/science.1167742},
    URL = {https://www.science.org/doi/abs/10.1126/science.1167742},
    eprint = {https://www.science.org/doi/pdf/10.1126/science.1167742}
}

@article{FloridiLea20_NatureScopesLimitsConseqsGPT3, 
    title={GPT-3: Its Nature, Scope, Limits, and Consequences}, 
    volume={30}, 
    ISSN={1572-8641}, 
    url={https://doi.org/10.1007/s11023-020-09548-1}, 
    DOI={10.1007/s11023-020-09548-1}, 
    number={4}, 
    journal={Minds and Machines}, 
    author={Floridi, Luciano and Chiriatti, Massimo}, 
    year={2020}, 
    month=dec, 
    pages={681–694}, 
    language={en}
}

@misc{HendrycksDea23p_OverviewCatastrophicAI, 
    title={An Overview of Catastrophic AI Risks}, 
    url={http://arxiv.org/abs/2306.12001}, 
    DOI={10.48550/arXiv.2306.12001}, 
    note={arXiv:2306.12001 [cs]}, 
    number={arXiv:2306.12001}, 
    publisher={arXiv}, 
    author={Hendrycks, Dan and Mazeika, Mantas and Woodside, Thomas}, 
    year={2023a}, 
    month=oct
}

@inproceedings{WeidingerLea22_TaxonomyRisksPosed, 
    address={New York, NY, USA}, 
    series={FAccT ’22}, 
    title={Taxonomy of Risks posed by Language Models}, 
    ISBN={978-1-4503-9352-2}, 
    url={https://dl.acm.org/doi/10.1145/3531146.3533088}, 
    DOI={10.1145/3531146.3533088}, 
    booktitle={Proceedings of the 2022 ACM Conference on Fairness, Accountability, and Transparency}, 
    publisher={Association for Computing Machinery}, 
    author={Weidinger, Laura and Uesato, Jonathan and Rauh, Maribeth and Griffin, Conor and Huang, Po-Sen and Mellor, John and Glaese, Amelia and Cheng, Myra and Balle, Borja and Kasirzadeh, Atoosa and Biles, Courtney and Brown, Sasha and Kenton, Zac and Hawkins, Will and Stepleton, Tom and Birhane, Abeba and Hendricks, Lisa Anne and Rimell, Laura and Isaac, William and Haas, Julia and Legassick, Sean and Irving, Geoffrey and Gabriel, Iason}, 
    year={2022b}, 
    month=jun, 
    pages={214–229}, 
    collection={FAccT ’22}
}

@inproceedings{BenderEea21_LLMsAsStochasticParrots, 
    address={New York, NY, USA}, 
    series={FAccT ’21}, 
    title={On the Dangers of Stochastic Parrots: Can Language Models Be Too Big?}, 
    ISBN={978-1-4503-8309-7}, 
    url={https://dl.acm.org/doi/10.1145/3442188.3445922}, 
    DOI={10.1145/3442188.3445922}, 
    booktitle={Proceedings of the 2021 ACM Conference on Fairness, Accountability, and Transparency}, 
    publisher={Association for Computing Machinery}, 
    author={Bender, Emily M. and Gebru, Timnit and McMillan-Major, Angelina and Shmitchell, Shmargaret}, 
    year={2021}, 
    month=mar, 
    pages={610–623}, 
    collection={FAccT ’21}
}

@article{degroot1974reaching,
  title={Reaching a consensus},
  author={DeGroot, Morris H},
  journal={Journal of the American Statistical association},
  volume={69},
  number={345},
  pages={118--121},
  year={1974},
  publisher={Taylor \& Francis}
}

@article{park2024generative,
  title={Generative agent simulations of 1,000 people},
  author={Park, Joon Sung and Zou, Carolyn Q and Shaw, Aaron and Hill, Benjamin Mako and Cai, Carrie and Morris, Meredith Ringel and Willer, Robb and Liang, Percy and Bernstein, Michael S},
  journal={arXiv preprint arXiv:2411.10109},
  year={2024}
}

@inproceedings{fontana2025nicer,
  title={Nicer Than Humans: How Do Large Language Models Behave in the Prisoner's Dilemma?},
  author={Fontana, Nicol{\'o} and Pierri, Francesco and Aiello, Luca Maria},
  booktitle={Proceedings of the International AAAI Conference on Web and Social Media},
  volume={19},
  pages={522--535},
  year={2025}
}

@article{corso2025early,
  title={Early linguistic fingerprints of online users who engage with conspiracy communities},
  author={Corso, Francesco and Russo, Giuseppe and Pierri, Francesco and Morales, Gianmarco De Francisci},
  journal={arXiv preprint arXiv:2506.05086},
  year={2025}
}

@inproceedings{sun2023opinion,
  title={Opinion optimization in directed social networks},
  author={Sun, Haoxin and Zhang, Zhongzhi},
  booktitle={Proceedings of the AAAI Conference on Artificial Intelligence},
  volume={37},
  pages={4623--4632},
  year={2023}
}

@article{disaro2024balancing,
  title={Balancing homophily and prejudices in opinion dynamics: An extended Friedkin--Johnsen model},
  author={Disar{\`o}, Giorgia and Valcher, Maria Elena},
  journal={Automatica},
  volume={166},
  pages={111711},
  year={2024},
  publisher={Elsevier}
}

@article{HegselmannRea02_OpinionDynamicsBounded,
  title = {Opinion {{Dynamics}} and {{Bounded Confidence Models}}, {{Analysis}} and {{Simulation}}},
  author = {Hegselmann, Rainer and Krause, Ulrich},
  year = {2002},
  month = jun,
  journal = {Journal of Artificial Societies and Social Simulation},
  volume = {5},
  number = {3},
  publisher = {{Journal of Artificial Societies and Social Simulation}},
  urldate = {2026-01-07}
}

@article{corso2025androids,
  title={Do Androids Dream of Unseen Puppeteers? Probing for a Conspiracy Mindset in Large Language Models},
  author={Corso, Francesco and Pierri, Francesco and Morales, Gianmarco De Francisci},
  journal={arXiv preprint arXiv:2511.03699},
  year={2025}
}

@article{nogara2026longitudinal,
  title={A longitudinal analysis of misinformation, polarization and toxicity on Bluesky after its public launch},
  author={Nogara, Gianluca and Sahneh, Erfan Samieyan and DeVerna, Matthew R and Liu, Nick and Luceri, Luca and Menczer, Filippo and Pierri, Francesco and Giordano, Silvia},
  journal={Online Social Networks and Media},
  volume={51},
  pages={100342},
  year={2026},
  publisher={Elsevier}
}

@inproceedings{di2022vaccineu,
  title={VaccinEU: COVID-19 vaccine conversations on Twitter in French, German and Italian},
  author={Di Giovanni, Marco and Pierri, Francesco and Torres-Lugo, Christopher and Brambilla, Marco},
  booktitle={Proceedings of the International AAAI Conference on Web and Social Media},
  volume={16},
  pages={1236--1244},
  year={2022}
}

@inproceedings{pierri2023political,
  title={Political advertisement on Facebook and Instagram in the run up to 2022 Italian general election},
  author={Pierri, Francesco},
  booktitle={Proceedings of the 15th ACM Web science conference 2023},
  pages={13--22},
  year={2023}
}

\section*{Paper Checklist}

\begin{enumerate}

\item For most authors...
\begin{enumerate}
    \item  Would answering this research question advance science without violating social contracts, such as violating privacy norms, perpetuating unfair profiling, exacerbating the socio-economic divide, or implying disrespect to societies or cultures?
    \answerYes{Yes.}
  \item Do your main claims in the abstract and introduction accurately reflect the paper's contributions and scope?
    \answerYes{Yes.}
   \item Do you clarify how the proposed methodological approach is appropriate for the claims made? 
    \answerYes{Yes, in ``Experimental Design''.}
   \item Do you clarify what are possible artifacts in the data used, given population-specific distributions?
    \answerYes{Yes, we discuss biases and limitations in ``Discussion and Conclusion''}
  \item Did you describe the limitations of your work?
    \answerYes{Yes, limitation are presented and discussed in ``Discussion and Conclusion''}
  \item Did you discuss any potential negative societal impacts of your work?
    \answerYes{We discuss negative societal impact in ``Ethical Statement''.}
      \item Did you discuss any potential misuse of your work?
    \answerYes{Yes, we discuss potential misuse in ``Ethical Statement''.}
    \item Did you describe steps taken to prevent or mitigate potential negative outcomes of the research, such as data and model documentation, data anonymization, responsible release, access control, and the reproducibility of findings?
    \answerYes{Yes, we share models' versions in ``Experimental Design'' and the full prompts in the Appendix}
  \item Have you read the ethics review guidelines and ensured that your paper conforms to them?
    \answerYes{Yes.}
\end{enumerate}

\item Additionally, if your study involves hypotheses testing...
\begin{enumerate}
  \item Did you clearly state the assumptions underlying all theoretical results?
    \answerNA{NA}
  \item Have you provided justifications for all theoretical results?
    \answerNA{NA}
  \item Did you discuss competing hypotheses or theories that might challenge or complement your theoretical results?
    \answerNA{NA.}
  \item Have you considered alternative mechanisms or explanations that might account for the same outcomes observed in your study?
    \answerNA{NA}
  \item Did you address potential biases or limitations in your theoretical framework?
    \answerNA{NA}
  \item Have you related your theoretical results to the existing literature in social science?
    \answerNA{NA}
  \item Did you discuss the implications of your theoretical results for policy, practice, or further research in the social science domain?
    \answerNA{NA}
\end{enumerate}

\item Additionally, if you are including theoretical proofs...
\begin{enumerate}
  \item Did you state the full set of assumptions of all theoretical results?
    \answerNA{NA.}
	\item Did you include complete proofs of all theoretical results?
    \answerNA{NA.}
\end{enumerate}

\item Additionally, if you ran machine learning experiments...
\begin{enumerate}
  \item Did you include the code, data, and instructions needed to reproduce the main experimental results (either in the supplemental material or as a URL)?
    \answerYes{Yes. The prompts are shared in the Appendix}
  \item Did you specify all the training details (e.g., data splits, hyperparameters, how they were chosen)?
    \answerNA{NA}
     \item Did you report error bars (e.g., with respect to the random seed after running experiments multiple times)?
    \answerYes{Yes, we report 95\% confidence intervals.}
	\item Did you include the total amount of compute and the type of resources used (e.g., type of GPUs, internal cluster, or cloud provider)?
    \answerYes{Yes, we use models hosted on a Cineca server and run on Ollama. We provide the cluster's specifications.}
     \item Do you justify how the proposed evaluation is sufficient and appropriate to the claims made? 
    \answerYes{Yes, specified in ``Experimental Design''}
     \item Do you discuss what is ``the cost`` of misclassification and fault (in)tolerance?
    \answerNA{NA.}
  
\end{enumerate}

\item Additionally, if you are using existing assets (e.g., code, data, models) or curating/releasing new assets, \textbf{without compromising anonymity}...
\begin{enumerate}
  \item If your work uses existing assets, did you cite the creators?
    \answerYes{Yes, we provide citations for all external assets we used.}
  \item Did you mention the license of the assets?
    \answerNA{NA}
  \item Did you include any new assets in the supplemental material or as a URL?
    \answerNA{NA}
  \item Did you discuss whether and how consent was obtained from people whose data you're using/curating?
    \answerNA{NA.}
  \item Did you discuss whether the data you are using/curating contains personally identifiable information or offensive content?
\answerNA{NA.}
\item If you are curating or releasing new datasets, did you discuss how you intend to make your datasets FAIR?
\answerNA{NA.}
\item If you are curating or releasing new datasets, did you create a Datasheet for the Dataset? 
\answerNA{NA.}
\end{enumerate}

\item Additionally, if you used crowdsourcing or conducted research with human subjects, \textbf{without compromising anonymity}...
\begin{enumerate}
  \item Did you include the full text of instructions given to participants and screenshots?
    \answerNA{NA.}
  \item Did you describe any potential participant risks, with mentions of Institutional Review Board (IRB) approvals?
    \answerNA{NA.}
  \item Did you include the estimated hourly wage paid to participants and the total amount spent on participant compensation?
    \answerNA{NA.}
   \item Did you discuss how data is stored, shared, and de-identified?
   \answerNA{NA.}
\end{enumerate}

\end{enumerate}

\appendix
\section{Appendix}
\label{sec:appendix}
\subsection{Computational Resources}
Using \llname{}, each run took approximately 8 hours (\(\sim\)1 hour for 3 simulated days) on Cineca's Leonardo cluster, booster partition. Hardware specifications were as follows:\textbf{CPU/node:} 32$\times$ Intel Ice Lake Intel Xeon Platinum 8358, \textbf{GPU/node:} 4$\times$ NVIDIA Ampere100 custom, 64 GB, \textbf{RAM:} 512 GB DDR4 

\subsection{Code and Data Availability}\label{sec:availability}

All code and prompts to replicate the analyses are available on GitHub: \url{github.com/elisacomposta/YAnalysis}.

\begin{algorithm}[h!]
\caption{Simulation procedure for each virtual day}
\SetAlgoLined
\ForEach{hour in 24}{
    \tcp{1. Active user sampling}
    $A \gets$ SampleActiveUsers()\;
    
    \tcp{2. Active users' actions}
    \ForEach{user $u \in A$}{
        content $\gets$ GetContent($u$)\;
        
        decision $\gets$ Decision(profile($u$), content)\;
        
        \Switch{decision}{
            
            \uCase{follow/unfollow}{
                UpdateGraph($u$, decision)\;
            }
            \uCase{post}{
                PublishPost($u$)\;
            }
            \uCase{comment}{
                PublishComment($u$, content)\;
            }
            \uCase{like}{
                AddReaction($u$, content)\;
            }
        }
    }
}
\tcp{3. End-of-day opinion update}
\ForEach{active user $u$}{
    info $\gets$ RecapInfo($u$)\;
    
    stance $\gets$ UpdateOpinion(profile($u$), info)\;
    
    stance $\in [-1, +1]$\;
}
\end{algorithm}

\subsection{Dataset Information}
The ITA-ELECTION-2022 is a multi platform social media dataset described in~\citet{pierri2023ita}.
For the purpose of our study we employed only the Twitter portion of the dataset composed of 19,087,594 tweets posted by 618,089 unique accounts.
The data was collected with the Twitter API over the period of the General Italian Elections, from September to October 2022.
Political affiliation was inferred following the methodology proposed in~\cite{pierri2024drivers}. 
The approach relies on a set of 471 Twitter accounts belonging to elected members of the Italian Senate and Chamber of Deputies, grouped into four major political coalitions and provided as part of the ITA-ELECTION-2022 dataset. 
These accounts are used as labeled reference points to propagate political labels to ordinary users.
Specifically, users who retweeted at least one politician were assigned a political affiliation based on the coalition they retweeted most frequently. 
User-level toxicity was derived by aggregating the toxicity scores of tweets authored by each user.

\subsection{Prompts}
\label{app:prompts}
This section contains all the prompts used throughout this work to guide the behavior of LLM agents, including those for initialization, interaction, content generation, and opinion update.

\subsection{Agent roleplay}
\label{app:agent}
Before performing any action, agents are initialized with a detailed profile that defines their identity, including political orientation and current opinions, and provides them complete descriptions of the topics and the opinions held by their supported coalition.

\begin{tcolorbox}[prompt]
You are role-playing as \texttt{\{name\}}, a \texttt{\{age\}}-year-old \texttt{\{nationality\}} \texttt{\{gender\}}, and you only speak \texttt{\{language\}}. You are \texttt{\{oe\}}, \texttt{\{co\}}, \texttt{\{ex\}}, \texttt{\{ag\}}, and \texttt{\{ne\}}.

Current \texttt{\{nationality\}} political topics include: \texttt{\{topic\_descriptions\}}.

You politically identify as \texttt{\{leaning\}}. This party has historically promoted the following principles:\\
\texttt{\{coalition\_opinion\}}.

These principles have shaped your initial worldview and personal beliefs.

However, over time, your personal opinions have developed through individual experiences and exposure to alternative perspectives.\\
Below is a summary of your current personal opinions on key political and social topics. These may reflect, diverge from, or expand upon your party's stance:\\
\texttt{\{opinion\}}
\end{tcolorbox}

\subsection{Actions}
\label{app:prompt_actions}
The following are the prompts for the actions that agents can perform when they are active.
Please note that the prompts for \textit{post} and \textit{comment} refer to base agents, while those for misinformation agents are provided in the next subsection.

\noindent\textbf{Post}
\begin{tcolorbox}[prompt]
Write a tweet that discusses the following topic: \texttt{\{topic\}}.\\
 - Your tweet MUST be under 280 characters including spaces. If it exceeds this limit, the output is INVALID. Keep it short and sharp.\\
 - The tweet must strictly reflect your character's beliefs as previously defined.\\
 - Use an informal tone, appropriate for social media posts.\\
 - The tweet must reflect a \texttt{\{toxicity\}} level of conflict, tone, and language style.\\
  - Hashtags should be placed at the end.\\
 - Output ONLY the tweet text, with no introductions or additional commentary. Don't mention anything with '@'.
\end{tcolorbox}

\noindent\textbf{Comment}
\begin{tcolorbox}[prompt]
You are participating to a discussion about the following topic: \texttt{\{topic\}}. Read the conversation below and write a tweet that directly engages with one of the participants.
\smallskip
 - Your tweet MUST be under 280 characters including spaces. If it exceeds this limit, the output is INVALID. Keep it short and sharp.\\
 - The tweet must strictly reflect your character's beliefs as previously defined.\\
 - Use an informal tone, appropriate for social media posts.\\
 - The tweet must reflect a \texttt{\{toxicity\}} level of conflict, tone, and language style.\\
 - Begin with @username to address the user you are interacting with. Don't mention anything else with '@'.\\
 - Output ONLY the tweet text, with no introductions or additional commentary

\#\#CONVERSATION START\#\#

\texttt\{{conv\}}

\#\#CONVERSATION END\#\#
\end{tcolorbox}

\subsection{Opinion Update}
\begin{tcolorbox}[prompt]
You are updating your character's opinions based strictly on the interactions below. Be consistent with your character's beliefs and personality as previously defined.\\
- \texttt{\{bias\_instructions\}}\\
- Update only the following topics: \texttt{\{topics\}}\\
- Do not introduce external reasoning or general considerations.\\
- Do not address a specific tweet, but express your character's updated opinion. The opinion must reflect the character's position on the topic as defined in the topic descriptions, not their reaction to individual statements or posts.\\
- Don't mention anyone with '@'.\\
- Output EXACTLY one line per topic, following this structure:\\
\textless topic\textgreater: [\textless LABEL\textgreater] \textless thought\textgreater
 
\medskip
 
Where:\\
- \textless thought\textgreater must be a clear and concise sentence that reflects your current personal opinion.\\
- \textless LABEL\textgreater must be one of: [STRONGLY SUPPORTIVE], [SUPPORTIVE], [NEUTRAL], [OPPOSED], [STRONGLY OPPOSED]. Choose the label based on the direction and intensity of your character's past behavior and beliefs.\\
\hspace{1cm} - [STRONGLY SUPPORTIVE] or [STRONGLY OPPOSED]: the character holds a firm, clearly defined position with strong consistency over time and no indication of moderation.\\
\hspace{1cm} - [SUPPORTIVE] or [OPPOSED]: the character tends toward a position but with some openness or nuance.\\
\hspace{1cm} - [NEUTRAL]: the character's behavior or prior stance shows ambiguity, balance, or lack of clear positioning.\\
- DO NOT include additional formatting between topics.
 
 \medskip
 
 \#\#OUTPUT FORMAT STRUCTURE\#\#
 
 \smallskip
 \textless topic1\textgreater: [\textless LABEL\textgreater] \textless thought\textgreater\\
 \textless topic2\textgreater: [\textless LABEL\textgreater] \textless thought\textgreater\\...
 
 \smallskip
 \#\#END OF OUTPUT FORMAT STRUCTURE\#\#
 
 \medskip
 
 \#\#INTERACTIONS START\#\#
 
 \medskip
 \texttt{\{memory\}}
 
 \medskip
 \#\#INTERACTIONS END\#\#
\end{tcolorbox}

\subsection{Coalition opinions}
\label{app:coalition_opinions}
The following are the opinions of the coalitions considered in this work.
They also serve as the initial opinions for the supporting agents.

\subsection{Centre-Left}

\begin{tcolorbox}[prompt]
\begin{itemize}
    \item \textbf{Civil rights}:
        [STRONGLY SUPPORTIVE] Support for equal marriage and adoption rights for same-sex couples, anti-homotransphobia laws, and recognition of LGBTQIA+ rights. 
    \item \textbf{Immigration}:
        [SUPPORTIVE] Policies of reception and inclusion are needed, aiming to facilitate integration pathways, guarantee migrants' rights, and build a European immigration management system based on solidarity among member states. Humanitarian corridors should be expanded for emergency situations.
    \item \textbf{Nuclear energy}: 
        [STRONGLY OPPOSED] The ecological transition must prioritize renewables and energy efficiency; nuclear power is considered too expensive, slow to implement, and incompatible with the urgent need to reduce emissions by 2030, while also raising unresolved environmental concerns.
    \item \textbf{Reddito di cittadinanza}
        [SUPPORTIVE] The current system shouldn't be abolished, but we should address distortions. Proposals include recalibrating the benefit, introducing support for large families, a minimum wage, mandating pay for curricular internships, and abolishing unpaid extracurricular internships.
\end{itemize}
\end{tcolorbox}

\subsection{Movimento 5 Stelle (M5S)}
\label{M5S_opinions}

\begin{tcolorbox}[prompt]
\begin{itemize}
    \item \textbf{Civil rights}:
        [STRONGLY SUPPORTIVE] Support for equal marriage, anti-homotransphobia legislation.
    \item \textbf{Immigration}:
        [SUPPORTIVE] A humanitarian approach is needed, with integration policies and mandatory redistribution of migrants across Europe.        
    \item \textbf{Nuclear energy}: 
        [STRONGLY OPPOSED] Nuclear energy has high costs and safety risks. We should focus on a decentralized energy model that encourages self-production and local energy efficiency.
    \item \textbf{Reddito di cittadinanza}
        [STRONGLY SUPPORTIVE] The reddito di cittadinanza is strongly defended, with proposals to enhance the efficiency of active labor policies and implement antifraud monitoring mechanisms.
\end{itemize}
\end{tcolorbox}

\subsection{Right}
\label{Right_opinions}
\begin{tcolorbox}[prompt]
\begin{itemize}
    \item \textbf{Civil rights}:
        [STRONGLY OPPOSED] We should avoid reforms introducing new rights regarding family and gender identity, with a preference for defending the 'traditional family.'
    \item \textbf{Immigration}:
        [STRONGLY OPPOSED] We should stop illegal immigration, with the support for stricter control policies, naval blockades, and flow management through bilateral agreements with countries of origin. We should create European-managed centers outside Europe to process asylum requests and distribute refugees fairly.
    \item \textbf{Nuclear energy}: 
        [STRONGLY SUPPORTIVE] We should support the development of next-generation nuclear power. This includes investment in research, production facilities, and integration with renewable energy sources to ensure energy security and reduce dependence on imports.
    \item \textbf{Reddito di cittadinanza}
        [STRONGLY OPPOSED] We should abolish the reddito di cittadinanza, with a preference for targeted support measures for employment and vulnerable groups to prevent abuse.
\end{itemize}
\end{tcolorbox}

\subsection{Third Pole}
\label{Third_Pole_opinions}

\begin{tcolorbox}[prompt]
\begin{itemize}
    \item \textbf{Civil rights}:
        [SUPPORTIVE] We need the introduction of laws against homophobia and transphobia, the creation of an Anti-Discrimination Authority.
    \item \textbf{Immigration}:
        [SUPPORTIVE] A regulated and planned immigration system is needed, with integration policies, regularization for those with jobs, and training pathways. Expanding humanitarian corridors and establishing a Ministry for Migration are also supported.       
    \item \textbf{Nuclear energy}: 
        [SUPPORTIVE] Including nuclear energy in the energy mix is needed to achieve the 'net zero emissions' goal by 2050, considering it necessary to meet future energy needs safely and efficiently.    
    \item \textbf{Reddito di cittadinanza}
        [OPPOSED] The current system is considered ineffective. It should be reformed to be reserved only for those unfit for work. The benefit should be revoked after the first job refusal, and a time limit should be imposed: if no employment is found within two years, the amount is reduced.
\end{itemize}
\end{tcolorbox}

\subsection{Additional Plots}
\label{app:plots}

\subsection{Inter-Group Interactions Example}
Figure extending the analysis presented in Section~\ref{subsec:interactions}
\begin{figure}[h]
    \centering
    \includegraphics[width=\linewidth]{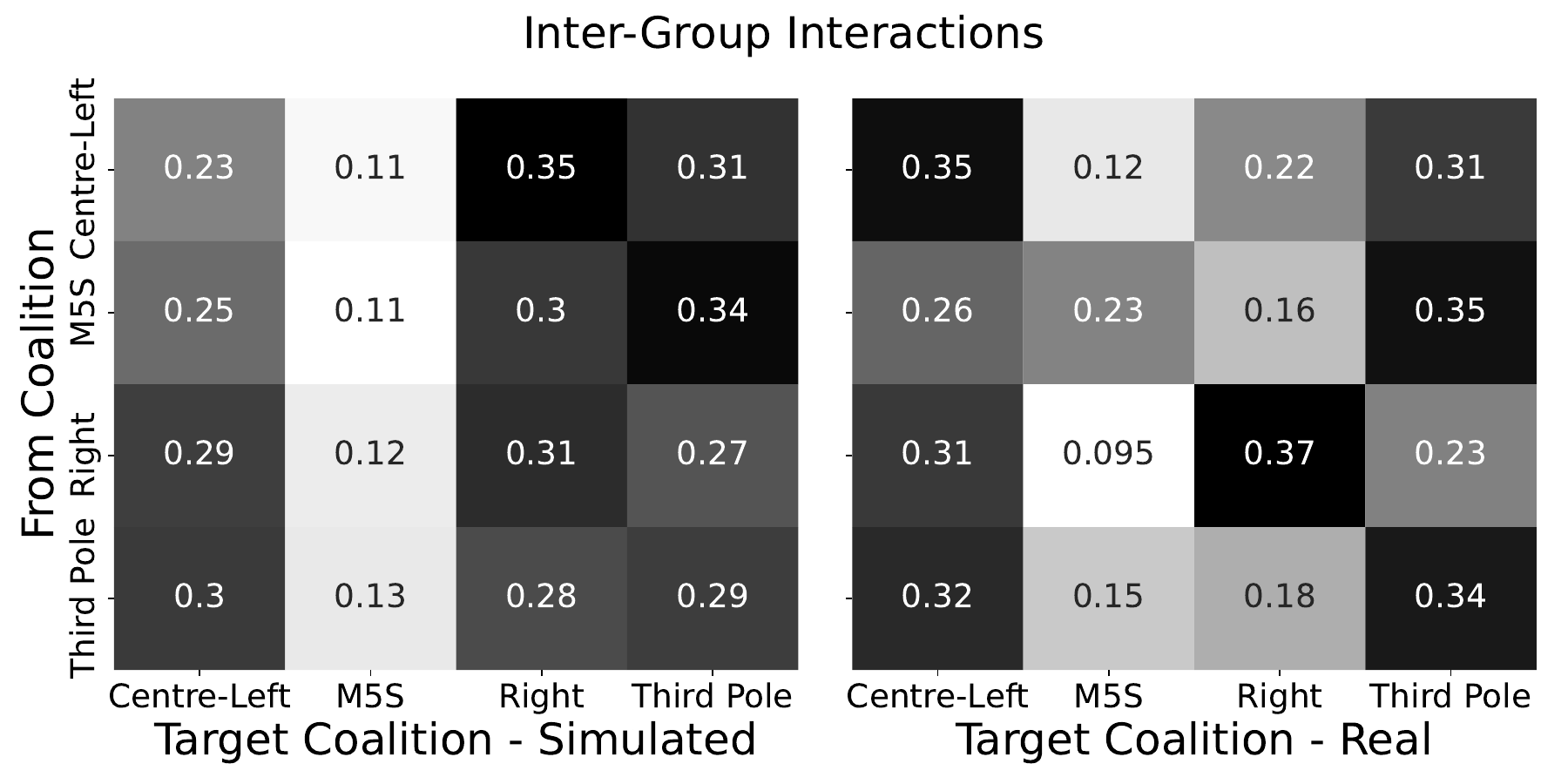}
    \caption{Example of comparison between simulated (10 runs across 1 configuration) and real data. On the left, the heatmap of inter-group interactions percentages for each coalition pair (using \artiname{}, empty initial network, and random recommender system). On the right, the same heatmap in the real-world dataset.}
    \label{fig:interactions_out_eg}
\end{figure}

\subsection{Inter-Group Toxicity Example}
Figure extending the analysis presented in Section~\ref{subsec:toxicity}
\begin{figure}[h]
    \centering
    \includegraphics[width=\linewidth]{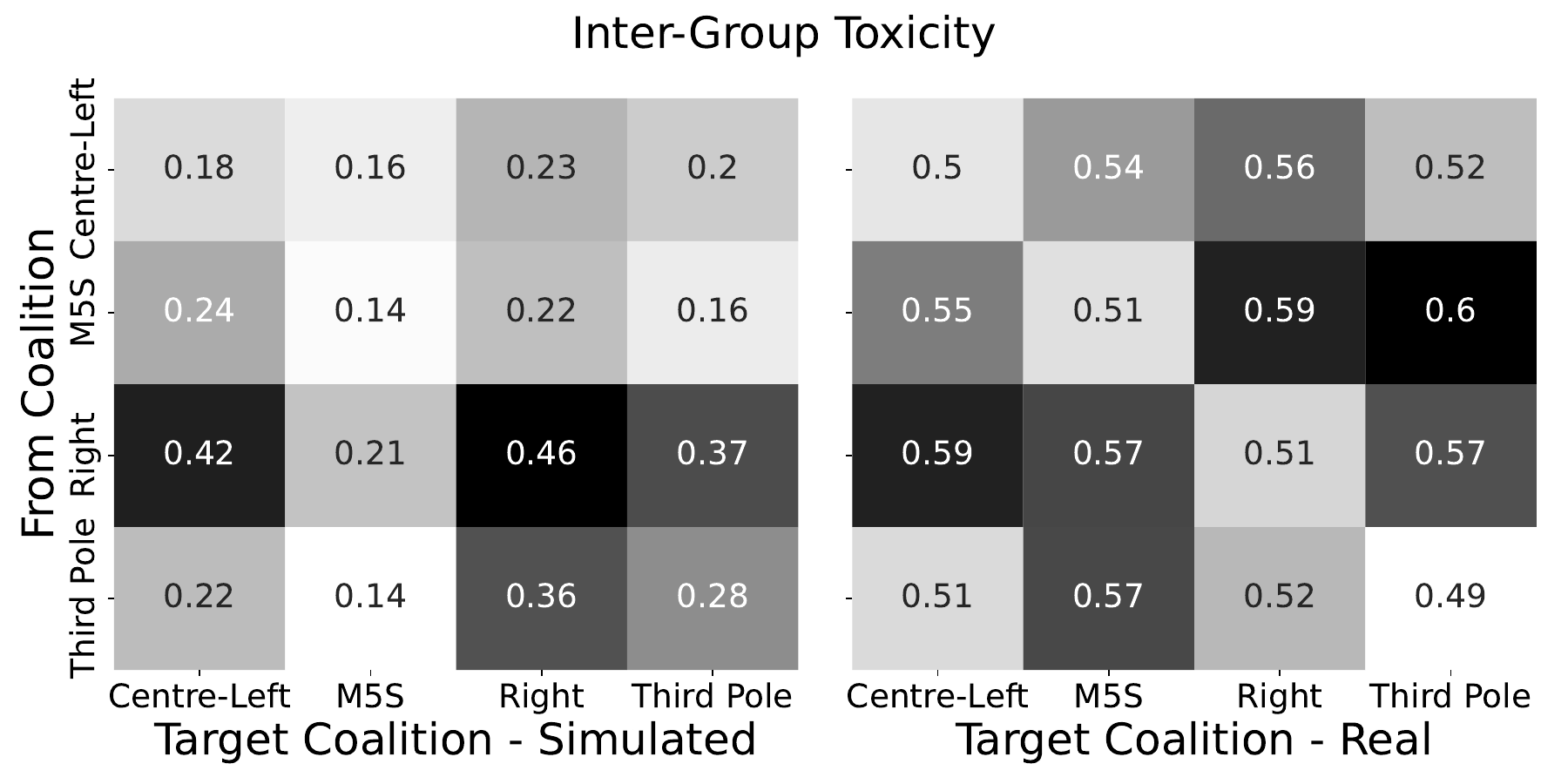}
    \caption{Example of comparison between simulated (10 runs across 1 configuration) and real data. On the left, the heatmap of inter-group toxicity percentages for each coalition pair (using \artiname{}, empty initial network, and random recommender system). On the right, the same heatmap in the real-world dataset.}
    \label{fig:toxicity_out_eg}
\end{figure}

\subsection{Opinion Shifts}
Figures extending the analysis presented in Section~\ref{subsec:opinion_dynamics}
\begin{figure}[h]
    \centering
    \includegraphics[width=\linewidth]{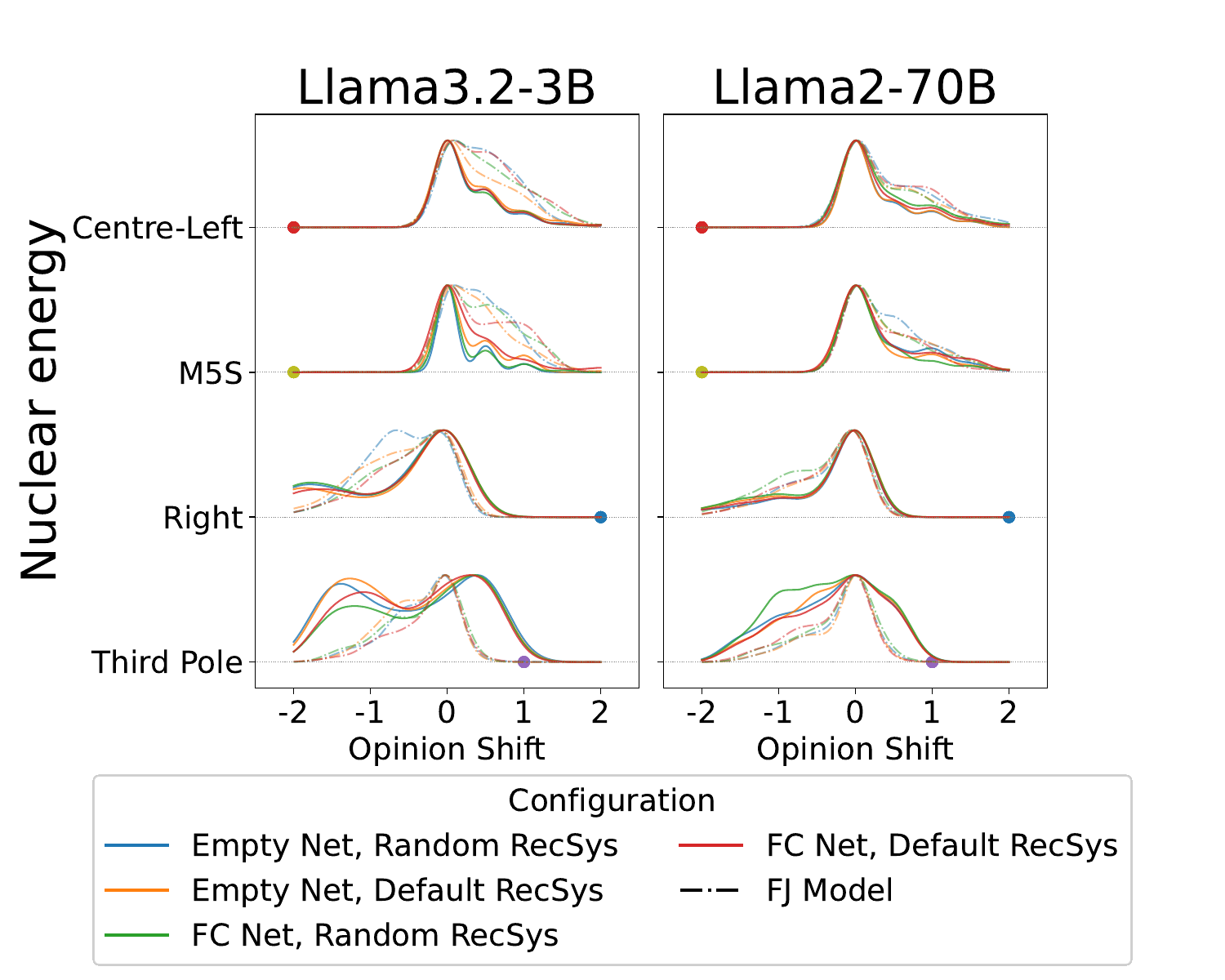}
    \caption{Topic: \texttt{Nuclear Energy}. Opinion shifts for each coalition for different configurations of model, network initialization, and recommender system. For each configuration, the corresponding simulation using the Friedkin–Johnsen mathematical model is reported (dashed lines).}
    \label{fig:opinion_shifts_nuclear}
\end{figure}
\begin{figure}[h]
    \centering
    \includegraphics[width=\linewidth]{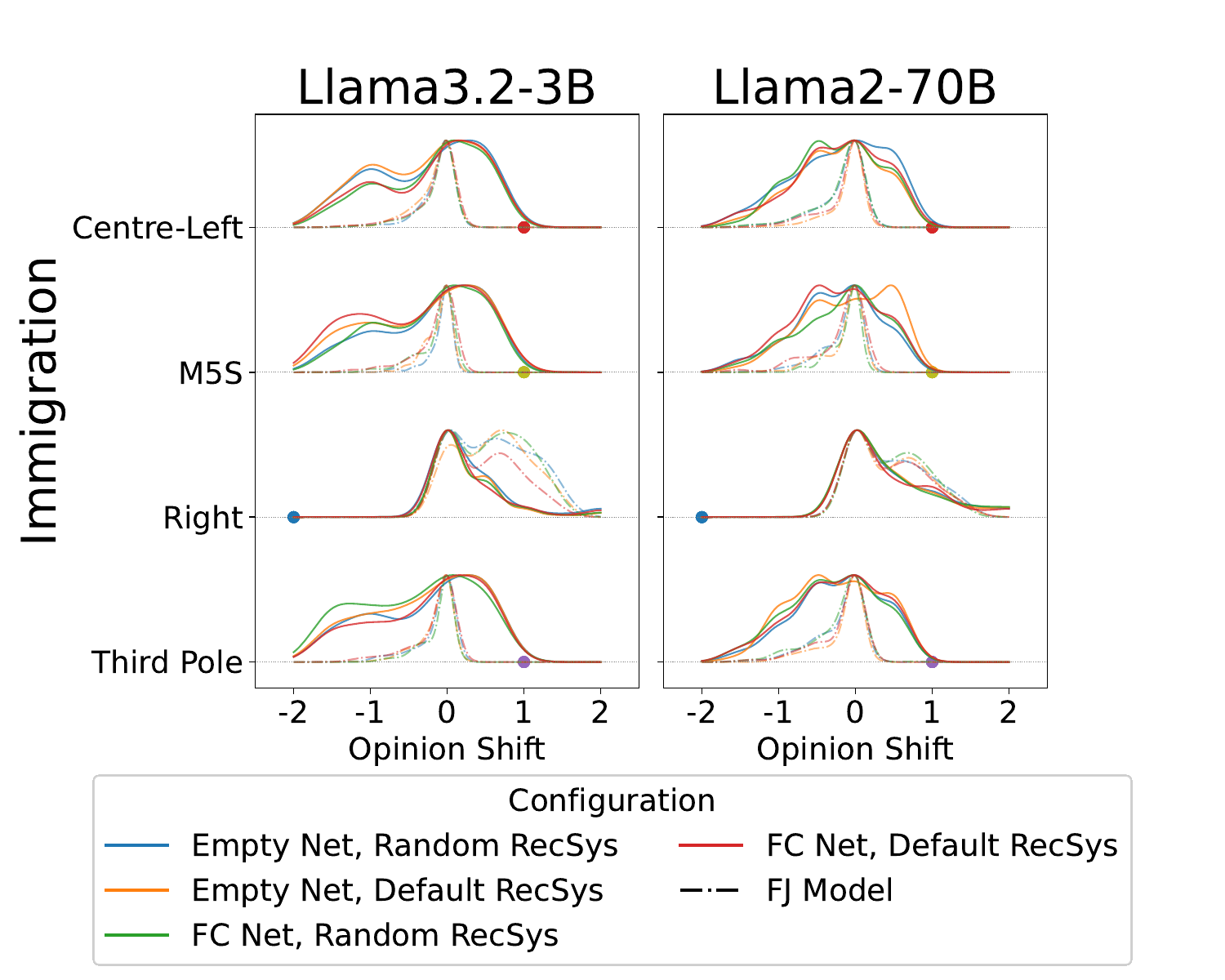}
    \caption{Topic: \texttt{Immigration}. Opinion shifts for each coalition for different configurations of model, network initialization, and recommender system. For each configuration, the corresponding simulation using the Friedkin–Johnsen mathematical model is reported (dashed lines).}
    \label{fig:opinion_shifts_immigration}
\end{figure}
\begin{figure}[h]
    \centering
    \includegraphics[width=\linewidth]{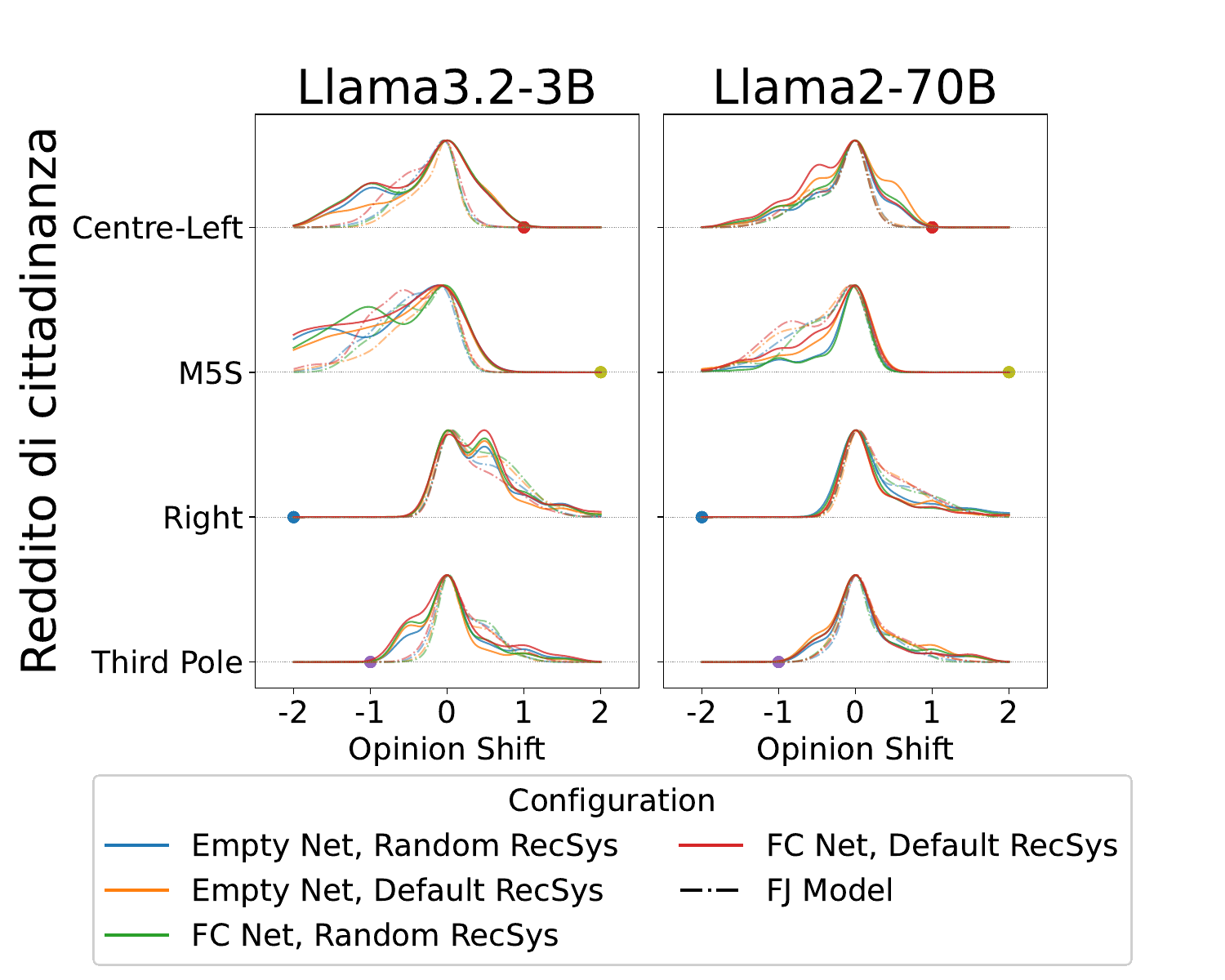}
    \caption{Topic: \texttt{Reddito di Cittadinanza}. Opinion shifts for each coalition for different configurations of model, network initialization, and recommender system. For each configuration, the corresponding simulation using the Friedkin–Johnsen mathematical model is reported (dashed lines).}
    \label{fig:opinion_shifts_reddito}
\end{figure}

\clearpage

\end{document}